\documentclass[12pt,preprint,dvips,deluxetable]{aastex}

\def \msun{$\mathrm{M}_\odot$}
\def \rsun{$\mathrm{R}_\odot$}

\def \kms{km~s$^{-1}$}
\def \1pap{Paper I}
\def \pap2{Paper II}
\def \deg{$^\circ$}

\makeatletter
\newcommand\cutinheadb[1]{%
 \noalign{\vskip .8ex}%
 \@ptabularcr
 \noalign{\vskip -4ex}%
 \multicolumn{\pt@ncol}{c}{#1}%
 \@ptabularcr
 \noalign{\vskip .8ex}%
 \hline
 \@ptabularcr
 \noalign{\vskip -1.5ex}%
}%
\def\cutinheadb@ppt#1{%
 \noalign{\vskip .8ex}%
 \@ptabularcr
 \noalign{\vskip -1.5ex}
 \multicolumn{\pt@ncol}{c}{#1}%
 \@ptabularcr
 \noalign{\vskip .8ex}%
 \hline
 \@ptabularcr
 \noalign{\vskip -1.5ex}%
}%
\makeatother

\makeatletter
\newcommand\cutinheadc[1]{%
 \noalign{\vskip .8ex}%
 \@ptabularcr
 \noalign{\vskip -4ex}%
 \multicolumn{\pt@ncol}{c}{#1}%
 \@ptabularcr
 \noalign{\vskip .8ex}%
 \hline
 \@ptabularcr
 \noalign{\vskip -1.5ex}%
}%
\def\cutinheadc@ppt#1{%
 \noalign{\vskip .8ex}%
 \@ptabularcr
 \noalign{\vskip -1.5ex}
 \multicolumn{\pt@ncol}{c}{#1}%
 \@ptabularcr
 \noalign{\vskip .8ex}%
 \hline
 \@ptabularcr
 \noalign{\vskip -1.5ex}%
}%
\makeatother

\author{Daniel~C.~Kiminki\altaffilmark{1}\altaffilmark{2} \&\
Henry~A.~Kobulnicky\altaffilmark{1}}

\altaffiltext{1}{Dept. of Physics \& Astronomy, University of Wyoming,
Laramie, WY 82070} 
\altaffiltext{2}{Dept. of Astronomy, University of Arizona,
Tucson, AZ 85721} 

\begin{document}

\title{An Updated Look at Binary Characteristics of Massive Stars in
the Cygnus OB2 Association}

\begin{abstract} This work provides a statistical analysis of the
massive star binary characteristics in the Cygnus~OB2 Association
using radial velocity information of 114 B3--O3 primary stars and
orbital properties for the 24 known binaries. We compare these data to
a series of Monte Carlo simulations to infer the intrinsic binary
fraction and distributions of mass ratios, periods, and
eccentricities. We model the distribution of mass ratio, log-period,
and eccentricity as power-laws and find best fitting indices of
$\alpha=0.1\pm 0.5$, $\beta= 0.2\pm0.4$, and $\gamma=-0.6\pm0.3$
respectively. These distributions indicate a preference for massive
companions, short periods, and low eccentricities.  Our analysis
indicates that the binary fraction of the cluster is $44\pm8$\%\ if
all binary systems are (artificially) assumed to have P$<$1000 days;
if the power-law period distribution is extrapolated to $10^4$ years,
a plausible upper limit for bound systems, the binary fraction is
$\sim$90$\pm 10$\%.  Of these binary (or higher order) systems,
$\sim$45\% will have companions close enough to interact during pre-
or post-main-sequence evolution (semi-major axis $\lesssim$4.7 AU).
The period distribution for $P<27$~d is not well reproduced by any
single power-law owing to an excess of systems with periods around
3--5~days (0.08--0.31~AU) and a relative shortage of systems
with periods around 7--14~days (0.14--0.62~AU). We explore
the idea that these longer-period systems evolved to produce the
observed excess of short-period systems.  The best fitting binary
parameters imply that secondaries generate, on average, $\sim$16\% of
the V-band light in young massive populations.  This means that
photometrically based distance measurements for young massive clusters
\& associations will be systematically low by $\sim$8\% (0.16 mag in
the distance modulus) if the luminous contributions of unresolved
secondaries are not taken into account.
\end{abstract}


\section{Introduction}
In clusters containing high concentrations of massive stars (B3 and
earlier), binary and multiple systems provide a virtual cornucopia of
information about the massive stars and the nature of their formation
process. In addition to the frequency at which they occur, we may
examine the distribution of several quasi-preserved orbital parameters
(e.g., period, mass ratio, and eccentricity) and gain insight into
many open questions surrounding massive stars, such as whether they
form in isolation or as part of a competing menagerie of protostars
\citep[for a review see][]{Zinnecker07}. Other issues include whether
there exists an evolutionary link between massive binary systems and
type Ib/c supernovae \citep{Kobulnicky07,Podsiadlowski92}, whether
there is a correlation between the massive binary fraction and the
stellar density of the parent cluster \citep{Sana08,Garcia01}, whether
massive binaries are responsible for the runaway OB stars
\citep{blaauw61}, and what role these stars play in short and long
$\gamma$-ray bursts \citep{fryer99}. Previous and current endeavors to
characterize the binary properties of massive stars include the
galactic O-star binary study of \citet{garmany80}, the Galactic O-star
speckle survey of \citet{maiz04}, the multi-cluster O-star binary
statistical analysis of \citet{Garcia01}, the VLT-FLAMES studies of
the Large and Small Magellanic Clouds \citep{Evans06}, the lucky
imaging O-star study of \citet{maiz10}, the adaptive optics O-star
studies of \citet{turner08}, \citet{Duch01}, and \citet{Sana10b}, the
HST Fine Guidance Sensor survey of Cyg~OB2 by \citet{Saida11}, and the
individual open cluster binary spectroscopic studies of
\citet{Sana09}, \citet{Mahy09}, \citet{Sana08}, \citet{Rauw04},
\citet{debeck06}, \citet{hillwig06}, and \citet{Merm95}. These studies
have found a broad range of binary fractions
\citep[0--63\%;][]{Sana10a} and widely varying characterizations of
the intrinsic orbital parameter distributions, including single power
law descriptions with various indices and multiple-component
descriptions requiring many free parameters. Reasons for the diverse,
and seemingly incompatible, results in the literature stem from the
varying methodologies and biases inherent in the surveys
themselves. The majority of massive binary surveys in the literature
include either a small sample of O stars within a given cluster or
larger samples assembled from multiple smaller surveys encompassing
different clusters \& associations \citep[e.g.,
][]{garmany80,Gies87,Evans06,Mason09}. The larger surveys have the
advantage of better statistics, but the smaller surveys have the
advantage of a more homogeneous sample (same formation environment and
similar ages).

The Cyg OB2 radial velocity survey (\citealt{Kiminki07},
\citealt{Kiminki08}, \citealt{Kiminki09}, \citealt{Kiminki2010}, and
\citealt{Kobulnicky2012}, in prep; Papers I, II, III, IV, \&\ VI),
takes advantage of both a large and homogeneous sample of OB stars
(114 massive stars) with radial velocity coverage of up to 12 years to
probe the binary characteristics of massive stars.
\citet[][KF07]{Kobulnicky07} analyzed the Cyg~OB2 radial velocity
data from 1999--2005 and compared the raw velocities with the
expectations of Monte Carlo simulations over a range of mass ratio and
orbital separation distributions and at varying binary
fractions.\footnote{We define the massive star binary fraction to mean
the number of systems with two (or more) components divided by the
total number systems containing at least one massive star. We caution
the reader that other definitions exist in the literature.  } They
find that the likely binary fraction of OB stars within Cyg~OB2 is
greater than 70\%\ for an \citet{Opik24} period distribution (i.e.,
log-flat) and mass ratio distribution best characterized by a
power-law slope of $\alpha=-0.6$ to $1.0$ (depending on the choice of
binary fraction). Though KF07 are very thorough in addressing
different possible distribution scenarios (a second ``twin''
population, a \citealt{Hogeveen92} distribution, a \citealt{Miller79}
secondary mass distribution, etc), they are forced to make certain
simplifying assumptions in their code based on the limited binary
information available. A few of the more significant assumptions are
circular orbits, treating all higher order systems as binaries rather
than triples, quadruples, etc, and finally, that all velocity
variation stems from orbital dynamics rather than stellar atmosphere
line profile variations, which may be present in massive
stars. Assuming circular orbits can cause an underestimation of the
binary fraction because fewer circular binaries need to be generated
to simulate the seemingly singular eccentric systems that exist in the
cluster. With regard to the next assumption, treating
(difficult-to-detect) triple and quadruple systems as binaries still
yields a meaningful estimate of the binary fraction of the cluster, as
long as this term is understood to include possible higher order
systems.  Lastly, the assumption that all velocity variation stems
from orbital dynamics can lead to an overestimation of the binary
fraction or an overestimation of number of massive companions, but
this is an unavoidable limitation to the KF07 analysis.

We present here a follow-up to the initial results of Papers I, II,
III, IV, \&\ VI, and the earlier analysis of KF07. We use the results
of nearly 12 years of spectroscopic observations obtained on 114
massive stars in the Cyg~OB2 Association to model the observed mass
ratio, orbital period, and eccentricity distributions composed from
the orbital information of 22 close massive binaries in the
cluster. We make use of all published close binary information for the
cluster, both photometric and spectroscopic, including single-lined
spectroscopic binaries (SB1s), double-lined spectroscopic binaries
(SB2s), and eclipsing binaries. We also use the radial velocity
information of 110 OB stars from our original sample from \citet{MT91}
to infer the binary fraction of the cluster. Additionally, we examine,
in more detail, an apparent excess of binary systems with periods
between 3 and 5~days that have been observed in Cyg~OB2
\citep{Kiminki09} and other clusters \citep[NGC~6231;][]{Sana08}. The
current work provides an analysis using the largest sample of massive
stars and close massive binary systems ever compiled for a group of
stars with a shared history.

Section~2 reviews the data used in this work. Section~3 summarizes the
method employed in this work to model the orbital period,
eccentricity, and mass ratio distributions using a Monte Carlo
approach. Section~4 presents the results of the Monte Carlo
simulations and discusses the most probable intrinsic orbital
parameter distributions and the probable binary fraction of massive
stars within the cluster. Discussion in Section~5 explores possible
physical origins for the observed orbital distributions and summarizes
results that may inform models of close binaries as progenitors to
supernova and $\gamma$-ray bursts.  Finally, Section~6 summarizes the
survey findings.

\section{The Data}
We obtained radial velocities for a total of 146 OB stars over 12
years. These 146 (given in Table~1 of Paper~I) were chosen from the
\textit{UBV} photometric and spectroscopic survey of Cyg~OB2 by
\citet{MT91}. Nearly half of the stars were previously identified as
an OB-type and the other half had colors consistent with
classification of B3 or earlier. In this work, we utilize 110 of the
146 \citet{MT91} stars, termed the ``unbiased'' sample because they
were selected on a ``binary-blind'' basis (i.e., no previous evidence
of a companion). The objects in the ``unbiased'' sample average 14
observations over 1999-2011 with a mean velocity uncertainty of
9.7~\kms.  Stars were removed from the original sample for having
fewer than three observations at sufficiently high S/N needed to
obtain reliable velocities (i.e., S/N$<$30 where velocity
uncertainties are $>$40 \kms) and for having spectral types later than
B3.  Additionally, we include the published orbital information for
four systems not in our original sample, but likely members of
Cyg~OB2: Schulte~5 from \citet{Schulte58}, A36, A45, and B17 from
\citet{Comeron02}. This constitutes 114 OB stars and is designated the
``complete'' sample. The primary components in the ``complete'' sample
have spectral types that range from B3V (e.g., MT298) to O3If (e.g.,
MT457) and masses that range from 7.6~\msun\ to 80~\msun\
\citep[initial masses were interpolated from the stellar evolutionary
models of][]{Lejeune2001} with a mean mass of $\sim$20~\msun\ and mean
temperature class of B0 (71 B3--B0 stars and 43 O9.5--O3 stars).

There are fewer stars than included in KF07 (who included stars from
Table~5 of Paper~I) owing to removal of seven stars having temperature
classes later than B3 (MT196, MT239, MT271, MT453, MT493, MT576, and
MT641), 13 stars having fewer than three observations with sufficient
S/N (MT238, MT343, MT435, MT477, MT490, MT492, MT493, MT509, MT568,
MT621, MT645, MT650, and MT712), and one star that is a newly
turned-on Be star (MT213).  Additionally, ten stars not included in
the analysis of KF07 but included here are Schulte~3, Schulte~73,
MT005, MT179, MT186, MT222, MT241, MT267, MT421, and MT426. These ten
stars were excluded from the earlier work because they had fewer than
three observations prior to 2007. In total, this survey has produced
full or partial orbital solutions for 19 binary systems. Combined with
the other published data, there are 24 known OB binary systems in
Cyg~OB2. The locations of the first 20 are shown by the open circles
in Figure~16 of Paper~IV and listed in Table~6 therein. The remaining
4 are presented in Paper~VI.

\section{Current Approach to Monte Carlo Analysis}
In keeping with prior works, we characterize the orbital parameter
distributions as single power-laws. The mass ratio distribution is
represented by $f(q)\propto q^\alpha$ (where $q=M_{2}/M_{1}$) over a
range of $q_{MIN}$ to 1, the period distribution is represented by
$f(\mathrm{log} P)\propto(\mathrm{log} P)^\beta$ over a range of
periods from $P_{MIN}$ to $P_{MAX}$~days, and the eccentricity
distribution is represented by $f(e)\propto e^\gamma$ over a range of
$e_{MIN}$ to $e_{MAX}$.  In the current approach, we constrain the
indices, $\alpha$, $\beta$, $\gamma$, by comparing the 12 years of
radial velocity data, in addition to measured mass ratios, orbital
periods, and eccentricities for the 22 binaries in Cyg~OB2 with
periods under 30~days, to Monte Carlo-generated binary populations.
For the standard application of the Monte Carlo (MC) code, we adopt
$q_{MIN}=0.005$ (roughly representing the mass ratio of a B3V~$+$~M8V
system; \citealt{drilling}), $P_{MIN}=1$~day, $P_{MAX}=1000$~days,
$e_{MIN}=0.0001$, and $e_{MAX}=0.9$.  The period upper limit is chosen
based on our observing campaign time span of a few thousand days. The
eccentricity upper limit is chosen based roughly on where an orbital
system becomes unbound. We show later that our results are not
sensitive to the exact choices for these upper and lower bounds, the
main exception being that the inferred binary fraction does depend on
the choice of $P_{MAX}$.

We created synthetic binary populations using Monte Carlo (MC) methods
by randomly drawing periods, mass ratios, and eccentricities from
power-law distributions with indices ranging from -2 to 2 in
increments of 0.1, with binary fractions ranging from 20\%--100\%\ in
increments of 5\%.  The MC code then generates a 4-dimensional
parameter space composed of $41\times41\times41\times17=1,171,657$
distinct ``families'' that are described by unique combinations of
each parameter.  During each iteration, radial velocities are
calculated for a group of 110/114 synthetic stars whose primary masses
are determined from the adopted values for Cyg~OB2 primaries.  The
number of synthetic velocities for each system is determined by the
actual number of observations for each given star in the survey.  The
code samples the synthetic velocity curve at intervals corresponding
to the actual observing cadence for that star.  If the system is, at
random, designated a binary, a mass ratio, eccentricity, and period
are drawn from the power-law distributions characterized by the family
currently being examined. An inclination is randomly drawn from a
probability distribution described by $\mathrm{sin}(i)$ over an
interval of 0 to $\pi/2$, representing a random orientation on the
unit sphere \citep{Law2009}.  The orbital phase and angle of
periastron are randomly drawn from uniform distributions over their
allowed values (i.e., 0--2$\pi$). The code adds Gaussian noise to each
simulated radial velocity, where the magnitude of the noise is set by
the observational uncertainty from the actual observations of the
selected star.  For example, when a system representing MT720 is
simulated, a primary mass of 17.5~\msun\ is used and 32 radial
velocities are calculated using the dates and errors listed in
Table~\ref{XX.tab}.  Finally, the code produces 300 realizations of
each family, designated a ``generation''. We compare the average
properties of each generation to the Cyg~OB2 data.

We constrain the binary fraction by comparing the $\chi^2$ probability
density function (PDF) for the simulated radial velocities to that of
the unbiased sample of Cyg~OB2 stars. We use the same method described
in Paper~I, where the probability corresponds to the likelihood that
the $\chi^2$ for each star's radial velocity measurements, as
considered around the weighted mean velocity, would be exceeded by
chance given $\nu=N_{OBS}-1$ degrees of freedom. We compare the
observed $\chi^2$ PDF to simulated ones for each generation using a
2-sided Kolgorov-Smirnov (K--S) test. In our tests of the code, we
were able to recover the binary fraction of a simulated dataset to
within $\pm5$\%.

In order to compare the Monte Carlo distributions of $q$, $P$, and $e$
with the observed ones, it was necessary to apply a series of filters
to the synthetic data to remove all binary systems with orbital
solutions undetectable by this survey (i.e., the systems are
designated ``single'' and the orbital parameters are not included in
the final distributions). We first eliminate all synthetic systems
have $\chi^2$ probabilities $>$5\%; such systems either have large
errors, few measurements, and/or small velocity amplitudes that
preclude full orbital solutions. The next filter removes all synthetic
systems that have a semi-amplitudes $K_{MIN}<$15~\kms, the minimum
semi-amplitude for which we can derive a complete orbital solution
given typical velocity uncertainties.  In exceptional cases where the
stellar lines are narrow, we are able to do better.  For example,
MT174, with $K_1=9$ \kms\ (Paper~VI), has the smallest velocity
semi-amplitude yet measured in the survey.  Finally, we retain only
systems that have periods within the 1.5--26.5~day range covered by
the 19 systems with orbital solutions where our survey is highly
complete (see Section~4.1).  The final mean extracted synthetic
orbital parameter distributions for each generation are then compared
to the observed ones using 2-sided K--S tests.  The final best-fit
combination of $\alpha$, $\beta$, $\gamma$ is determined by product of
the resultant individual K--S probabilities for each of the three
parameter distributions.

In our testing of the code, we produced a series of simulated datasets
(i.e., orbital parameter distributions and ephemerides), similar to
the Cyg~OB2 dataset, covering a large range of possible $\alpha$,
$\beta$, and $\gamma$ combinations and binary fractions. The power-law
indices ranged from $-2.5$ to $2.5$ (and the MC code search grid was
increased accordingly) and binary fractions ranged from 20\%\ (the
absolute minimum binary fraction of the Cyg~OB2 sample based on
current observations) to 100\%. The simulated datasets were
constructed using the observation cadences, velocity uncertainties,
and primary masses from the Cyg~OB2 data. The code was able to recover
the original distribution indices to within $\pm$0.3 rms and the
original binary fraction to within 5\% rms.  The magnitude of the
uncertainty on each index is limited by the current sample size of
N=22 systems having orbital solutions. The uncertainty on the binary
fraction is limited by the mean number of observations per object and
the mean observational uncertainty.

\subsection{Limitations in the Code, Dataset, and Knowledge of Cyg~OB2}
While our current approach explores a larger parameter space than
KF07, incorporates eccentric orbits, and utilizes the information of
22 close massive binary systems in addition to the radial velocity
information of 12 years of spectroscopic observations, there are still
several limitations stemming from our modeling approach, our dataset,
and our current knowledge of Cyg~OB2.

\noindent \textbf{Limitations owing to our modeling approach:} As in
KF07, we still consider all velocity variation as stemming from
orbital dynamics rather than atmospheric phenomena such as pulsations.
\citet{Clark2010} show that pulsations in evolved early-type stars can
significantly affect radial velocity measurements, meaning that this effect
may add scatter to radial velocity measurements in roughly one-quarter
of our sample. We also still treat higher order systems as
binaries. At present, we know of three triple or higher order systems,
Cyg~OB2~No.5, Cyg~OB2~No.8, and MT429 (see references in Table~6 of
Paper~IV). In addition, we limit characterization of each orbital
parameter distribution to a single power-law, an assumption that may
not hold.

\noindent \textbf{Limitations owing to the dataset:} As mentioned, the
Cyg~OB2 radial velocity survey probes binary systems with periods
ranging from one day to a few thousand days, meaning we are not
sensitive to extremely close (and extremely rare) systems with 
$P<$1 d.  
Therefore, the MC code only draws from period distributions with
periods ranging 1 to 1000~days. Additionally, the vast majority of
spectroscopic observations were obtained June through November when
Cygnus is visible to northern latitudes. This, coupled with typical
velocity uncertainties of 9.7~\kms, reduces our sensitivity to long
period, high eccentricity systems (i.e., systems that may pass through
periastron and exhibit their largest velocity variation during an
unobservable period), systems with extreme mass ratios, and systems
with small inclination angles. Even though we have detected and
computed solutions for systems with semi-amplitudes as small as
9~\kms\ (Paper~VI), we attempt to circumvent these limitations by
designating systems in the code with semi-amplitudes smaller than
$15$~\kms\ as ``single''.

Another limitation comes in the form of not being able to detect
unresolved SB2s. For systems exhibiting highly blended line profiles
(even at quadrature), such as long period binaries with mass and
luminosity ratios near unity, radial velocity variations of the
blended line profile will be small and possibly undetectable at the
resolution of the spectra in this survey, thereby being designated as
``single'' by the code. \citet{Sana09} discuss this effect in their
analysis of the massive star binary fraction of NGC~6611 and conclude
that the effect is minimal (affecting their modeling of the binary
fraction by only a few percent). Additionally, there is also a negligible 
bias toward detecting short-period, high-mass-ratio systems that show
up as SB2s as it only takes one observation at quadrature to determine
that the system warrants follow-up. These systems, however, have large
velocity separations (i.e., large semi-amplitudes) and will get
flagged as a ``binary'' by the code with only a few observations.
 
\noindent \textbf{Limitations owing to incomplete knowledge of
Cyg~OB2:} A few studies have shown that Cyg~OB2 may be contaminated by
an older population of stars, meaning that evolutionary effects may be
present \citep[e.g.,][]{Wright10,Drew08,Comeron08}. However,
\citet{Comeron08} point out that the field star contamination is
likely composed of F to late B stars at a radius of 1 degree. This
exceeds the radial extent of the region examined in this study and
concerns stars of a later type than are included here. On the other
hand, some doubt remains in the membership of the 114 OB stars
included in this work as recent findings by \citet{Stroud10} and
\citet{Linder09} provide conflicting distance estimates to Cyg OB2
based on their complete orbital solutions calculated for the eclipsing
SB2s, Cyg~OB2~No.5 ($\sim$1~kpc) and B17 (1.5--1.8~kpc)
respectively. We maintain that the contamination is minor, if it
exists, owing to the localized region we examine, but even the removal
of a few periods or mass ratios could influence the distributions we
observe at present. Similar to this is the assumption that the members
of Cyg~OB2 have a common age. It has been suggested that even within
the core region of Cyg~OB2, there has been more than one epoch of star
formation \citep{Wright10}. There are no clear predictions for how a
population of close binaries should evolve over the 1--5 Myr lifetime of
most massive stars in young clusters, 
so we have no way to model such effects or even
know if they are important.

\section{Results of the Monte Carlo Analysis}

We use the Monte Carlo code in conjunction with the ``complete''
sample of Cyg~OB2 data to infer the underlying distribution of orbital
parameters, that is, $\alpha$, $\beta$, $\gamma$.  The orbital
parameter information (but not ephemerides) for Schulte~9 and MT070
are excluded given their much longer periods.  In order to address the
binary fraction, however, we use the data of the ``unbiased''
sample. Ultimately, we find that the orbital parameter indices and
binary fraction are essentially identical for both the ``unbiased''
and ``complete'' samples.

\subsection{Period Distribution}
Binary surveys that examine the orbital period distribution probe
widely varying ranges in period/separation and stellar mass, and they
find widely varying characterizations of its functional form. For
solar-type stars with periods from 1--10,000+~days, \citet{DM91} find
a log-normal distribution.  Among massive stars, most studies find a
uniform distribution in log space ($\beta=0$), sometimes known as an
\citet{Opik24} distribution.  For example, \citet{garmany80} find
$\beta\sim0$ for a large number of Galactic O-star binaries in the
northern hemisphere. \citet{Kouwenhoven07} also find an \"{O}pik's
distribution for a survey of visual and spectroscopic binaries in
Sco~OB2 composed of A- and B-type primaries (for 0.7~days~$\le P\le
3\times10^8$~days). \citet{Zinnecker07} speculate that rather than a
simple power-law distribution for O-star binaries, there may be a
concentration of systems having periods between 2 and 5 days. The
O-star binaries of NGC~6231 show this trend \citep{Sana08}. However,
\citet{Sana10a} conclude that a broken \"{O}pik distribution provides
a better description of the period distribution for spectroscopic
O-binaries in nearby clusters, where the broken distribution could
better explain the higher concentration of systems with periods less
than 10~days.

The observed orbital period, eccentricity, and mass ratio
distributions for the 22 known close massive binaries in Cyg~OB2 are
shown in Figure~\ref{Disthist}. The orbital period and mass ratio
distributions from \citet{garmany80} are also shown and represented by
the dotted lines in the upper and lower panels, respectively. The
Cyg~OB2 and \citet{garmany80} period distributions show a
concentration between 3 and 5~days (with a peak around 4--5~days for
Cyg~OB2).  A possible explanation for such a concentration might be
observational bias owing to the average observing run lasting a few
days to a week. Three systems in Cyg~OB2 with periods between 3 and
5~days originate from other studies (MT421, MT429, and B17) where this
may be the case. However, the remaining five systems in this period
range originate from our study which has a time coverage of up to
12~years and individual observing run durations of 1--3 weeks. This
allows for a sensitivity to orbital periods up to 25~days without
appreciable bias. This abundance of 3--5 day systems is not limited to
the SB2s having the largest mass ratios either. Three of the nine SB1s
in Cyg~OB2 also have periods that are in the 4--5~days range. This is
illustrated in Figure~\ref{PEQ}, which shows eccentricity versus
orbital period (upper panel) and mass-ratio (lower panel) versus
orbital period. The SB1s and SB2s are the open and filled circles
respectively, and the asterisks represent the orbital periods and
eccentricities for the massive star binaries (primaries of type B3 and
earlier) in the Ninth Spectroscopic Binary Catalog \citep{Pour04}. One
can easily see the concentration of points around 3--5~days and a
peculiar shortage of systems with periods from 7--14~days in the
Cyg~OB2 data. An initial hypothesis for this shortage of points might
be that completeness plays a role. However, with multiple studies with
varying observing cadences contributing to the catalog of binaries in
Cyg~OB2, we maintain that a more likely observation would be to see a
continually diminishing number of binary detections with increasing
period as completeness declines. Given the existence of this excess in
periods around 3--5~days, we caution that a simple power-law may
be insufficient to describe the distribution of orbital periods for
early-type stars in this cluster.

In Figure~\ref{Biased1}, we show the results from the Monte Carlo
analysis of the 22 close OB binaries in the ``complete'' sample.  The
four panels display the $1\sigma$ (68.3\%), $2\sigma$ (95.4\%), and
$3\sigma$ (99.7\%) contours relative to the maximum for the average
probability of a generation as a function of two parameters for fixed
best-fit values of the remaining two parameters. For example, the
upper left panel shows the average probability as a function of
$\alpha$ and $\gamma$ for fixed best-fit values of $\beta$ (0.2) and
$B.F.$ (45\%). The upper right panel similarly shows the average
probability as a function of $\beta$ and $\gamma$ for fixed best-fit
values of $\alpha$ (-0.1) and $B.F.$ (45\%). The cross in each panel
marks the location of the best fit for each parameter (i.e., the
maximum of the average probability). We find that the period distribution
is confined at the 1$\sigma$ level to $-0.3<\beta<0.5$, with a best
fit of $\beta=0.2\pm0.4$.  This points to a nearly log-flat
distribution, consistent with \"{O}pik's Law \citep{Opik24},
 consistent with the previous
O-star findings of \citet{garmany80}, \citet{Kouwenhoven07},
\citet{Sana10a}, and KF07.  Figure~\ref{pdist} shows the cumulative
period distribution of the Cyg~OB2 data (diamonds) and the best fit
cumulative period distribution from the MC code (solid curve). The
hump in the distribution at $log(P)\sim 0.65$ or $P\sim 4.5$~days
greatly influences the fit shown, and it is clear that a power-law of
$\beta=0$ might be a better fit if this hump were not present or less
pronounced. Indeed, no single power-law provides a good fit to the
cumulative period distribution.  We address the peculiar nature of the
period distribution in Section~5.

Figure~\ref{completeness_per} shows the survey completeness as a
function of orbital period for the best-fitting power-law indices.  We
use the MC code to estimate the survey completeness by tallying the
fraction of synthetic systems deemed detectable in our survey given
the time sampling and velocity semi-amplitude limit of $K_{MIN}=15$
\kms.  The survey is 88\% complete at $P=10$ days and $\gtrsim83$\%\
complete at $P=30$~days.  Our completeness analysis implies that,
given 14 detected binaries with periods $P<10$ days, there are an
additional 1--2 undetected binary systems in this range.
Statistically, the Monte Carlo code shows that all of these would be
undetectable by the survey because they have some combination of low
inclination, small mass ratio, large eccentricity, and larger radial
velocity uncertainties.  If, for the sake of argument, we ignore the
excess of short-period systems and assume that the $\beta\sim0$
power-law holds, the $\sim$15 binaries in the 1--10 day period range
would imply an equivalent number between 10 and 100 days.  There are 8
known systems in this range, leaving an additional 7 putative binaries
as undetected.  The mean completeness over this range is $\sim82$\%,
suggesting that $\sim$12 of the 15 would be detectable by our survey.
This implies that the ongoing Cyg~OB2 survey should uncover as many as
$\sim$5 additional binaries with periods between 10 and 100 days,
given sufficient long-term data.  For periods longer than 100 days,
where there are presently only two systems with orbital solutions,
Schulte~9 \citep{Naze10} and MT070 (Paper~VI), detection becomes more
difficult.  In the case of Schulte~9, with a period of
$\sim$2.355~years and an eccentricity of nearly 0.7, even an observing
campaign such as this one experienced difficulty in uncovering this
massive ``twin'' system.  All of the foregoing discussion, however,
assumes an \"Opik period distribution, and we have already shown in
Figure~\ref{pdist} that no single power-law adequately reproduces the
data at the shortest periods, $P<$ 14 days.

\subsection{Eccentricity Distribution}
Observationally, binaries with solar-type primaries and periods longer
than a year have a broad distribution of orbital eccentricities with a
median value of 0.55 \citep{DM91}. For periods shorter than this,
these systems have a higher concentration near zero eccentricity
stemming from circularization of the orbits. This is also shown to be
true for O-star binaries with periods as long as 10,000~days
\citep{Sana10a}. Although the method by which binary orbits are
circularized and the critical period below which all are circular is
unclear from existing data, a standard eccentricity versus period
scatter diagram, like Figure~\ref{PEQ}, illustrates that binaries with
short periods do tend toward nearly circular orbits. The figure shows
a decline of mean eccentricity for periods under
10~days. \citet{Giuricin84} found that nearly all (92\%) early-type
binaries with periods shorter than two days had an eccentricity less
than 0.05.  They also found that the eccentricities of nearly all of
the early-type binaries in their survey could be explained by the
\citet{zahn77} tidal theory, which describes the circularization time
for an equal mass binary with an initial eccentricity of $e<0.3$ as
being related to the main sequence lifetime of that
binary. \citet{Khal10} provide a modified version of the
\citet{zahn77} theory and address the evolution of the orbital
parameters over the circularization time of the orbit. The
circularization timescales are, on average, a third of those
predicted by \citet{zahn77} and in better agreement than the
predictions of \citet{tassoul88} with the massive star binary
statistics of \citet{Giuricin84}. The predicted circularization time
for an early-type (pair of B0V stars), nearly equal-mass binary at a
separation comparable to the Cyg~OB2 early B-type binaries (e.g.,
$a\mathrm{sin}i\sim$0.16~AU for MT720) is close to the main-sequence
lifetime of an early B star. However, a study by \citet{witte01} found
that there are a significant number of early-type binaries with
periods up to 10~days with an eccentricity near zero that cannot be
explained with the \citet{zahn77} theory and are unlikely to have been
primordially circular. \citet{KrumThom07} have proposed a method by
which orbits become significantly circularized during the star
formation process, thereby diminishing the conflict between
circularization time and the theoretical life span of massive stars.
However, the theory assumes the binary is formed concurrently with the
formation of the components, and that is a highly debated topic in
massive star formation theory \citep{Zinnecker07}.

In Figure~\ref{edist} we show the MC code best fit to the Cyg~OB2
eccentricity distribution. As in Figure~\ref{pdist}, diamonds show the
normalized cumulative distribution (left Y-axis) and the solid curve
shows the normalized cumulative distribution for $\gamma=-0.6$.
Figure~\ref{Biased1} (upper left and right panels) shows that the
index $\gamma$ is confined, at the $1\sigma$ level, to
$-1.0<\gamma<-0.4$ with a best fit of $\gamma=-0.6\pm0.3$. The steep
negative power-law is required to reproduce the abundance of systems
with low eccentricity.  A K--S test shows that the $\gamma=-0.6$ power
law is compatible (i.e., the observed sample and the synthetic sample
are consistent with being drawn from the same parent distribution)
with the data at the $\sim$90\% level.  The absence of eccentricities
with $e>0.55$ is similar to what is observed for short-period binaries
in other OB clusters \citep[e.g. NGC~6231;][]{Sana08}. Highly
eccentric systems are apparently rare in this period regime, either
because circularization mechanisms are efficient, or possibly because
they are more easily disrupted during encounters with other cluster
stars.  Finally, we note that the eccentricity distribution within
Cyg~OB2 is also compatible with that of OB binaries in the 9th
Spectroscopic Binary Catalog at the $\sim$50\% level.

We show an estimate of the survey completeness as a function of
eccentricity for two different period ranges in
Figure~\ref{completeness_ecc}. The solid and dot-dash curves represent
the completeness for $P<1000$~days and $P<30$~days respectively. The
plot indicates that the survey is $\sim$75\%\ complete at $e\sim0.65$
for binaries with $P<1000$~days and $\sim$88\%\ complete at
$e\sim0.75$ for binaries with $P<30$~days.  For $P<30$ days, the
survey is likely to have detected the vast majority of systems, even
those with large eccentricities, if they exist.

\subsection{Mass Ratio Distribution}
The nature of the mass ratio distribution is one of the more highly
debated and unresolved questions pertaining to massive binaries and
their formation. The distribution is difficult to characterize because
of the unavoidable bias of observational studies, including this one,
against low-$q$ systems.  The general consensus is that short-period
binaries have a different mass ratio distribution compared to the
long-period binaries. \citet{goldberg91} and \citet{DM91} show this
with lower-mass binaries, but they also point out that the period
range over which this change occurs is still ill-defined. Most studies
on the mass ratio distribution are for mid to late-type stars. A few
examples include \citet{DM91}, who find $\alpha\le-2$ (for $q>0.1$ and
periods 1--10,000+~days) for a large survey of solar-type
spectroscopic binaries, \citet{Kouwenhoven07}, who find $\alpha=-0.4$
to $-0.5$ for a large survey of spectroscopic and visual A- and B-type
primaries in Sco OB2, \citet{Fisher05}, who find $\alpha\sim0$ for all
the SB1s and $\alpha>0$ for all the SB2s in the solar neighborhood
(for periods 1--500~days), and \citet{Hogeveen92}, who find
$\alpha\simeq0$ for mass ratios of $q<0.3$ and $\alpha\le-2$ for mass
ratios of $q\gtrsim0.3$ in binaries of B-type and later (for periods
$<$1 day to 10,000+~days). Among the massive stars, studies such as
\citet{pinson06} conclude from their study of 21 eclipsing massive
binaries ($P<5$~days) in the Small Magellanic Cloud, that the
secondary mass function for massive binaries does not follow a
\citet{Salpeter55} slope. Instead, the authors find a flat
distribution of mass ratios and predict that roughly 55\%\ of all
massive systems are part of a ``twin'' system with a mass ratio of
$q>0.95$. \citet{Sana10a} have refuted this proposition, but also
find a preference for more massive companions among O-stars in
clusters.

\begin{figure}
\centering
\epsscale{1.0}
\plotone{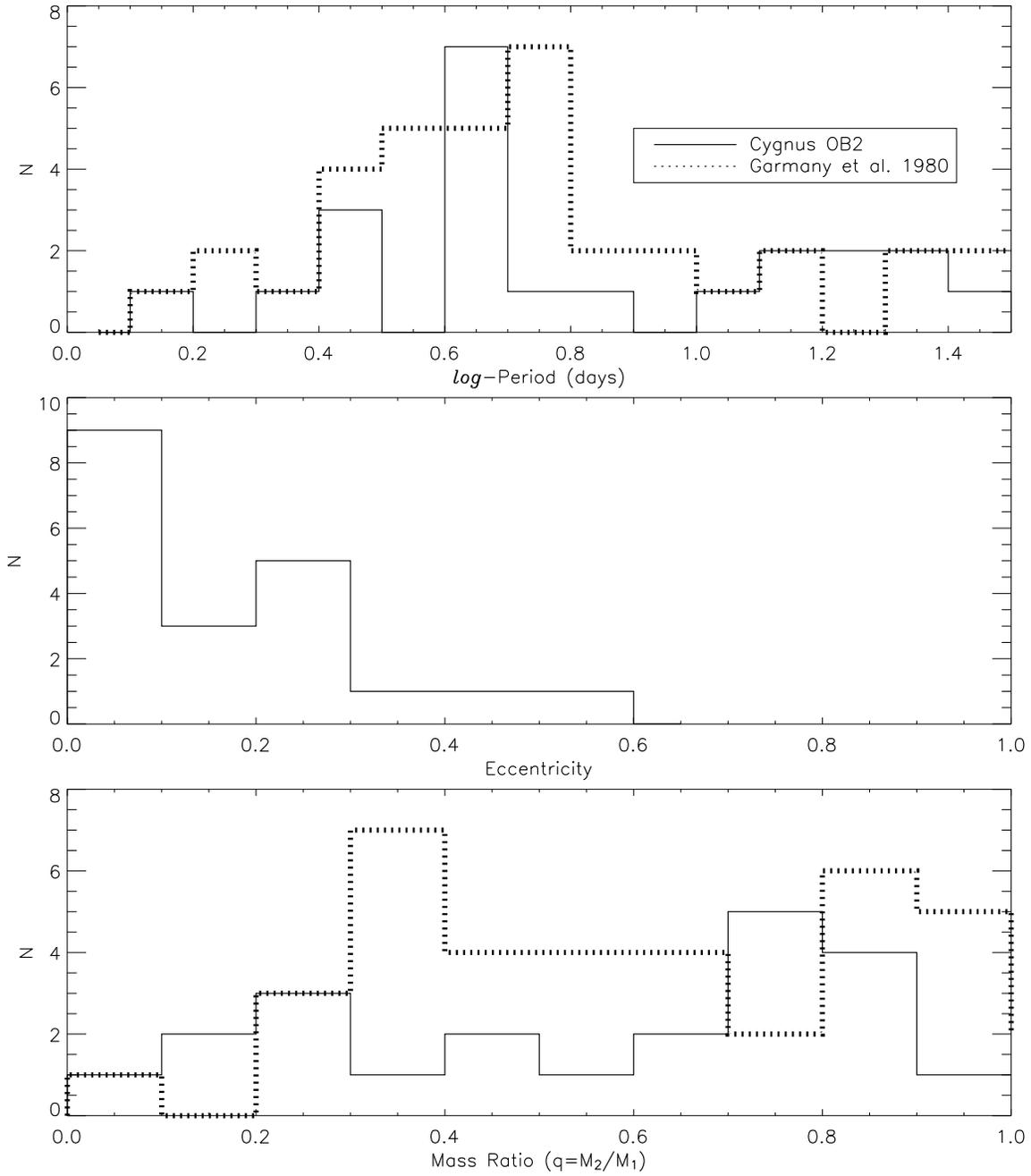}
\caption{Histograms of the Cyg~OB2 orbital period distribution in
$log$-Period (days) (top panel), orbital eccentricity distribution (middle
panel), and mass ratio distribution (bottom panel). The orbital period
and mass ratio distributions from \citet{garmany80} are illustrated by
the dashed lines.
\label{Disthist}}
\end{figure}

\begin{figure}
\centering
\epsscale{1.0}
\plotone{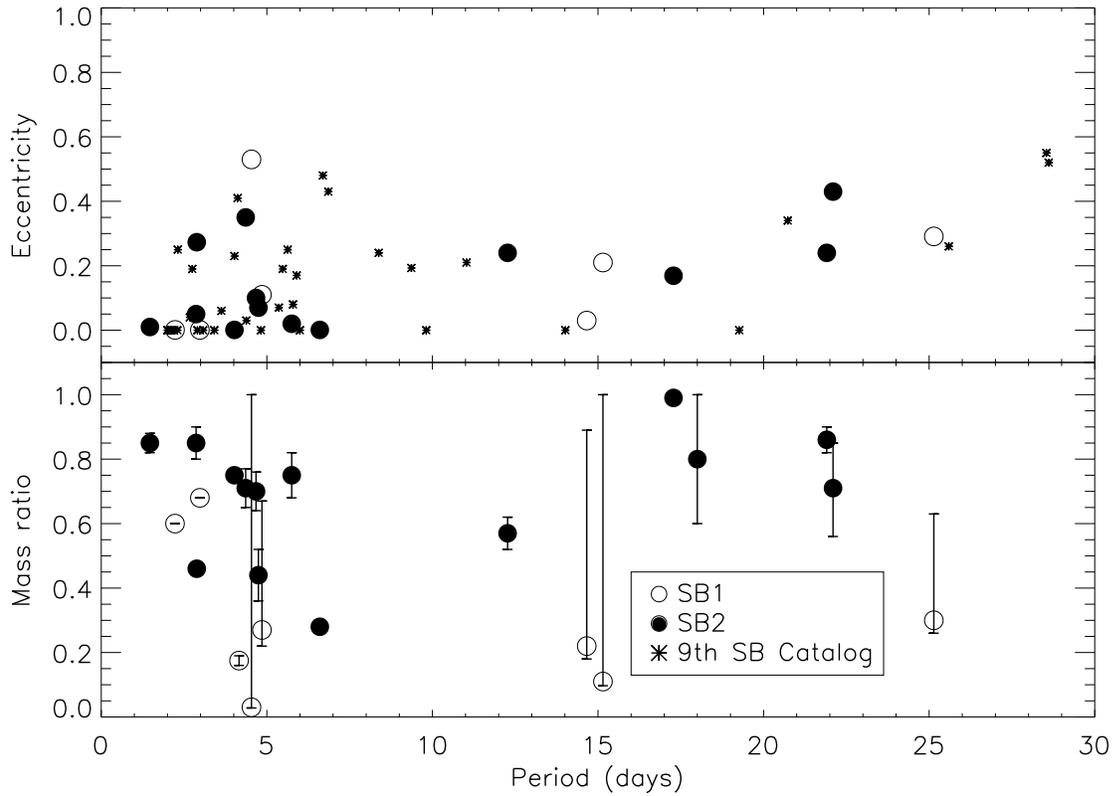}
\caption{Eccentricity versus period (top panel) and mass
ratio versus period (bottom panel) for 22 of the 24 known massive star
binaries in Cyg~OB2. SB1s are represented by open circles and SB2s are
represented by filled circles. The data 
for massive star binaries of the Ninth Spectroscopic Binary Catalog
\citep{Pour04} are over-plotted for comparison (asterisks).
\label{PEQ}}
\end{figure}

The distribution of observed mass ratios in Cyg~OB2, shown in
Figure~\ref{Disthist}, appears relatively flat, except for a slight
preference for mass ratios between 0.7 and 0.9. This may be an
artifact of using a standard histogram for small numbers. Not
surprisingly, the upper end of the distribution is populated
exclusively by SB2s, while the lower end of the distribution is almost
entirely populated by SB1s. For five of these single-lined systems, we
calculated the expected orbital inclination, $\langle i \rangle$,
between $i_{MIN}$ and 90\deg, using the inclination lower limits
estimated from Paper~II, Paper~III, and Paper~VI. We calculate the
most probable mass ratios to be 0.27, 0.30, 0.03, 0.22, and 0.11 for
MT059, MT145, MT174, MT258, and MT267, respectively. These are close
to the lower-limit values for $q$ because $\langle i \rangle$ is not
far from the upper limit on $i$.  There is a possibility, of course,
that for one or more of these systems, the orbital inclination could
be less than $\langle i \rangle$, and therefore, the system(s) would
have a more massive companion.  In such a case, our present estimates
may incorrectly inflate the number of systems with mass ratios between
0.05 and 0.30, and the true distribution would not be as flat as
currently shown in Figure~\ref{Disthist}.

From the MC code we find a best fit of $\alpha=0.1\pm0.5$ for the mass
ratio distribution.  Figure~\ref{qdist} shows the normalized
cumulative distribution of mass ratios (diamonds), along with the
corresponding simulated cumulative distribution for $\alpha=0.1$
(thick curve).  The 1$\sigma$ contours (top left and bottom left
panels in Figure~\ref{Biased1}) show a range of
$-0.3<\alpha<0.7$. Such a shallow positive power-law indicates that
mass ratios in the cluster are quasi-uniformly distributed between
$q=0.005$ and $q=1.0$ with a preference for larger mass ratios. This
type of distribution implies that the secondaries are not drawn from a
\citet{Salpeter55} stellar IMF, in contrast to lower-mass binary
populations which exhibit an abundance of low mass ratios.
Furthermore, the histogram of mass ratios in Figure~\ref{qdist} is
also inconsistent with the ``twin-heavy'' population suggested by
\citet{pinson06}, whereby 45\% of systems are predicated to have
$q>0.95$.

Figure~\ref{completeness_mrs} shows the completeness of the survey as
a function of mass ratio.  For systems with $P<30$~days (dot-dash
curve), the survey is at least 90\% complete down to $q\sim0.25$, but
there is a sharp drop in the fraction of systems detected for
$q\lesssim0.25$.  The average completeness is 10--20\% lower, overall,
for longer period systems with $P<1000$ days (solid curve).  Our
measurement of MT174 ($q=0.03$; $K_1=9$ \kms) below our nominal
 semi-amplitude
detection threshold of 15~\kms\ provides evidence that such
low-amplitude, low-$q$ systems exist and are detectable in our survey.
Figure~\ref{completeness_mrs} shows that incompleteness is an issue
only at the lowest mass ratios.  Any attempt to correct for
incompleteness at $q<0.25$ would add low-$q$ systems to the histogram
and serve to flatten the distribution.

\subsection{Binary Fraction}

Table~\ref{Ostat.tab} shows a selection of the larger surveys that
investigated the O-star binary fraction within individual
clusters/associations. Column~1 contains the common cluster
identification, column~2 is the number of O stars included in the
survey, column~3 is the number of early B stars included in the
survey, column~4 is the lower limit on the O-star binary fraction,
column~5 is the lower limit on the massive star binary fraction (O and
early-B primaries), and column~6 references the original works. The
observed O-star binary fractions (designated $b.f.$ rather than $B.F.$
here to distinguish from the intrinsic binary fraction) within the clusters
listed range widely from 0~--~67\% (column 4).  More inclusively, the
massive star $b.f.$ (with primaries B3 and earlier;
column~5), varies between 4\%\ (NGC~330 in the SMC) and 53\%\
(NGC~6231). In Paper I, we find an initial lower limit on the binary
fraction of 30\%\ based on the $\chi^2$ probability density function
for radial velocities from the first six years of observations of
Cyg~OB2, which would place it near the middle of the binary fraction
range among massive stars. KF07 find a more probable binary fraction
of $B.F.>70$\% based on their Monte Carlo analysis of the raw radial
velocities, which would place it closer to that of NGC~6231. However,
based purely on the binary detections of Papers~II--IV \&\ VI, the
hard lower limit on the binary fraction among massive stars in Cyg~OB2
is 21\% (i.e., 24/114).

We show the observed $\chi^2$ probability density function (PDF) for
the radial velocities of Cyg~OB2 stars in Figure~\ref{probchi}. There
are 53 having $P(\chi^2,\nu)<$5\%\ and 43 with $P(\chi^2,\nu)<$1\%.
These systems encompass the 24 known binary systems and additional
candidates for radial velocity variables. The number of variables has
increased from Paper~I despite removing some low S/N spectra that may
have caused extraneous velocities.  This is understandable given the
additional data acquired on the sample, especially on long period
binaries like MT070 ($P=7.58$~years). Figure~\ref{chisqdist} shows the
MC code best fit to the cumulative $\chi^2$ PDF for radial velocities
in Cyg~OB2 using the ``unbiased'' sample of 110 stars. The solid curve
represents the cumulative $\chi^2$ PDF produced by the MC code with
best-fit indices of $\alpha=0.1$ ($q_{MIN}$=0.005), $\beta=0.4$
($P_{MAX}$=1000~days), and $\gamma=-0.6$ ($e_{MIN}$=0.0001,
$e_{MAX}$=0.9). The open diamonds represent the normalized cumulative
$\chi^2$ PDF for the unbiased sample of OB primaries. The best MC code
fit results in a binary fraction of $B.F.=45\pm8$\% if all binary
systems are assumed to have $P<1000$ days.  Simulated binary
populations with binary fractions of 40--50\%\, in increments of 1\%
and with fixed $\alpha$, $\beta$, and $\gamma$, were individually
examined for better agreement with the data. These additional
simulations show that a binary fraction of 44\%\ yields a slightly
better fit to the data.

\begin{figure}
\centering
\epsscale{1.0}
\plotone{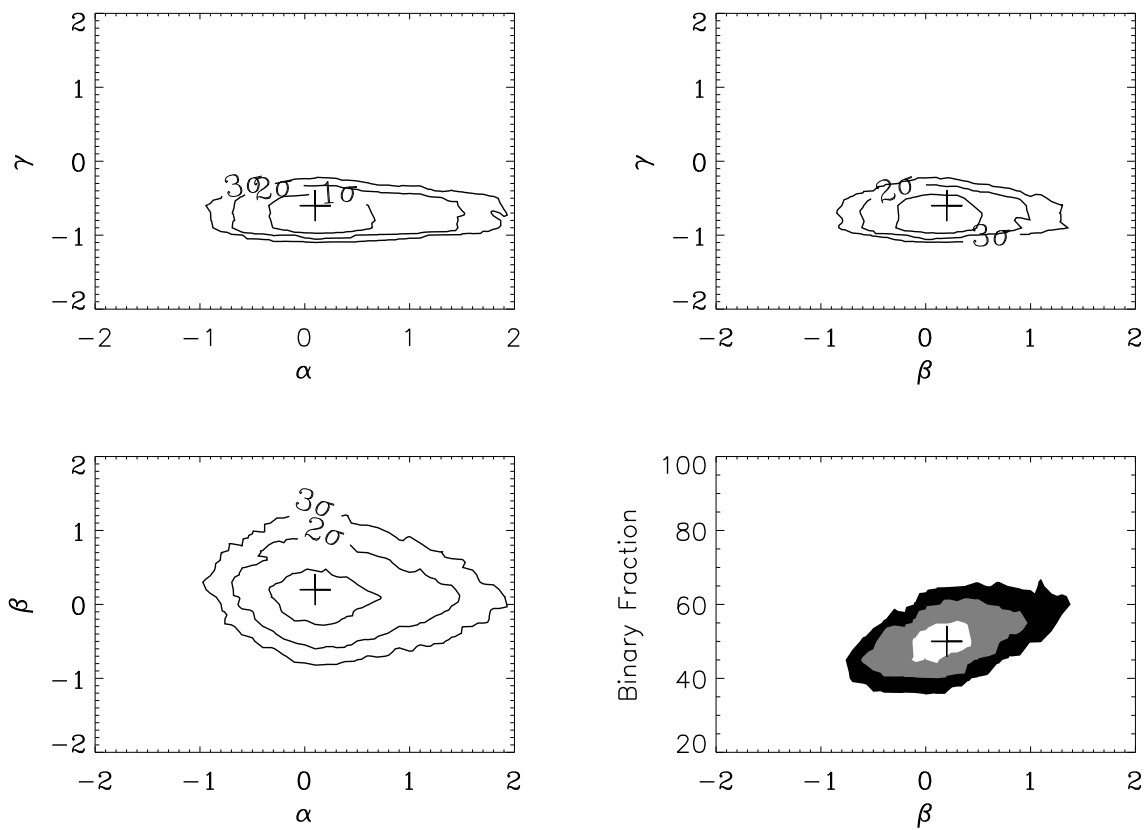}
\caption{Contour plots depicting the probabilities of K--S tests
between the observed orbital parameter and radial velocity data of the
``complete'' sample and the MC code fits. Contours in the upper left
panel are for a $\beta=0.2$ and $B.F.=50$\%. Contours in the upper right
panel are for a $\alpha=0.1$ and $B.F.=50$\%. Contours in the bottom left
panel are for a $\gamma=-0.6$ and $B.F.=50$\%. Contours in the bottom
right panel are for a $\alpha=0.1$ and $\gamma=-0.6$.
\label{Biased1}}
\end{figure}

\begin{figure}
\centering
\epsscale{1.0}
\plotone{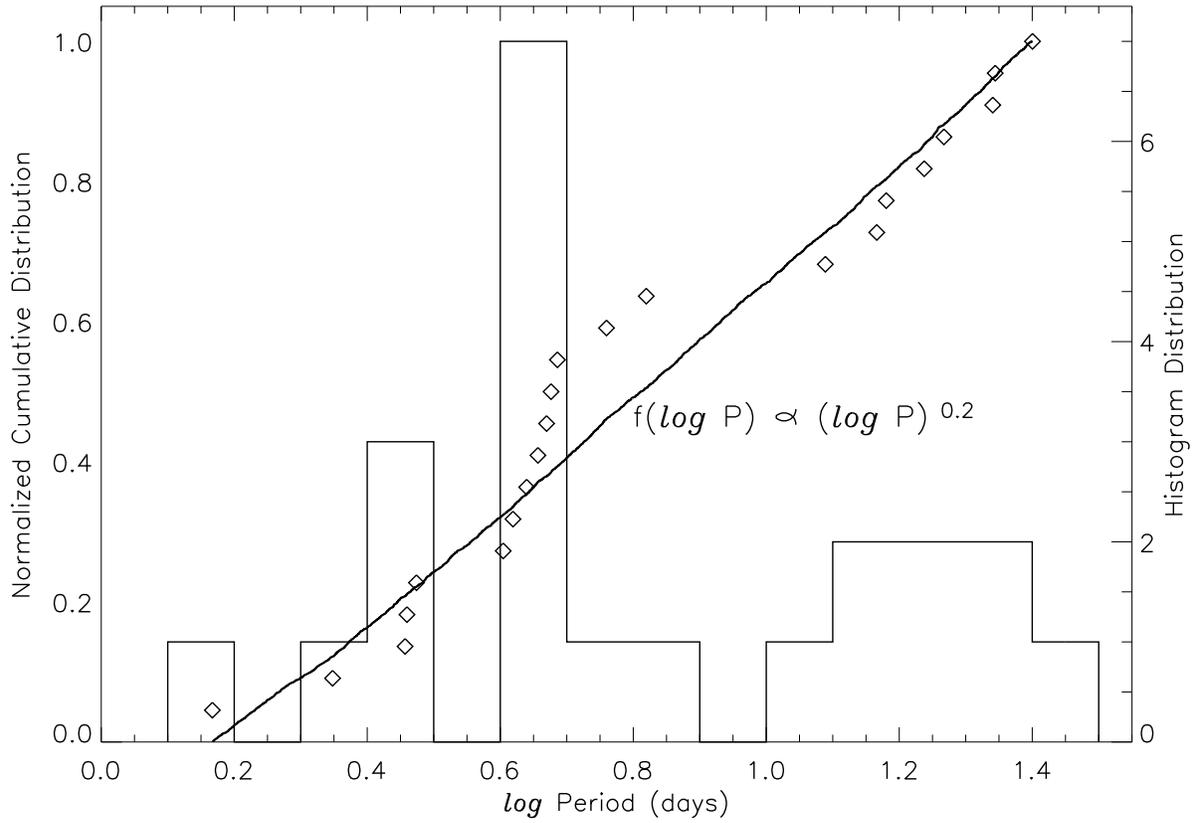}
\caption{Cumulative period distribution (diamonds) and 
conventional histogram for the
observed orbital periods. The solid curve is the cumulative period
distribution for the best MC code fit ($\beta=0.2$).  
\label{pdist}}
\end{figure}

\begin{figure}
\centering
\epsscale{1.0}
\plotone{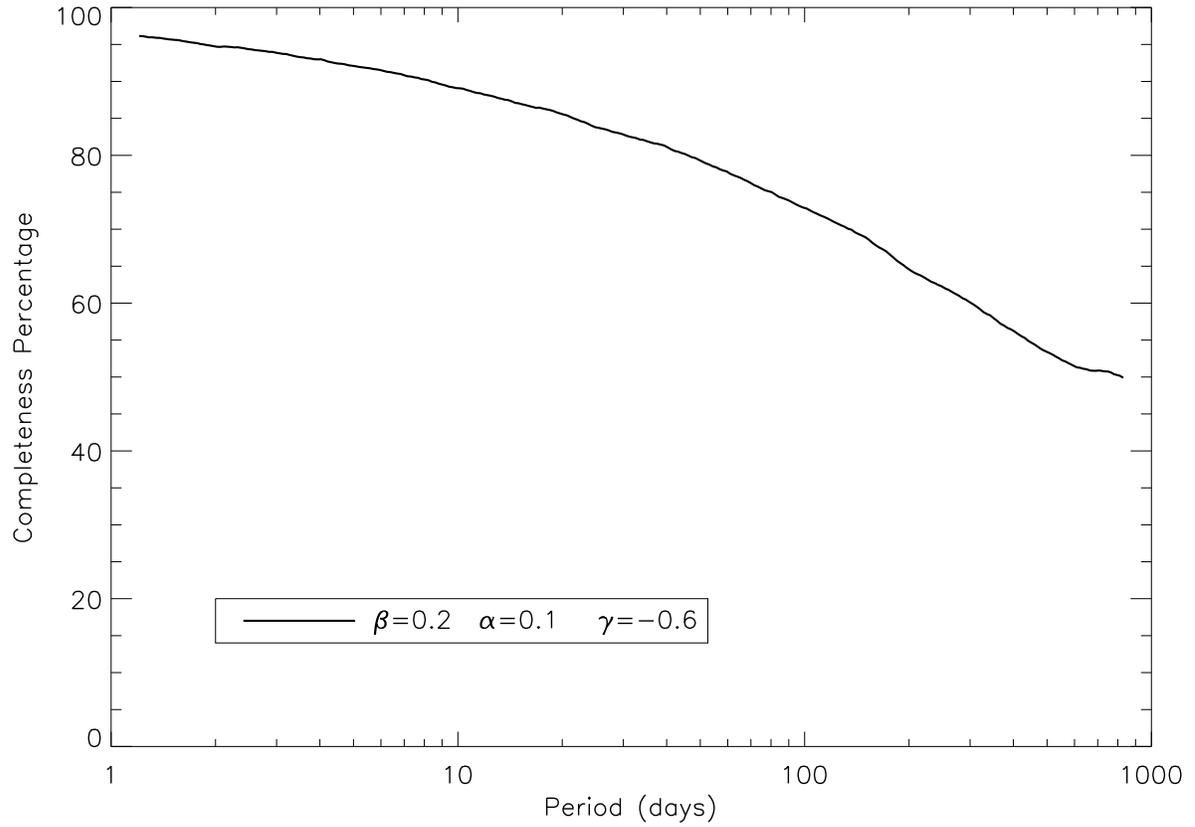}
\caption{Survey completeness as a function of orbital period for the best-fit 
power-law indices.
\label{completeness_per}}
\end{figure}

\begin{figure}
\centering
\epsscale{1.0}
\plotone{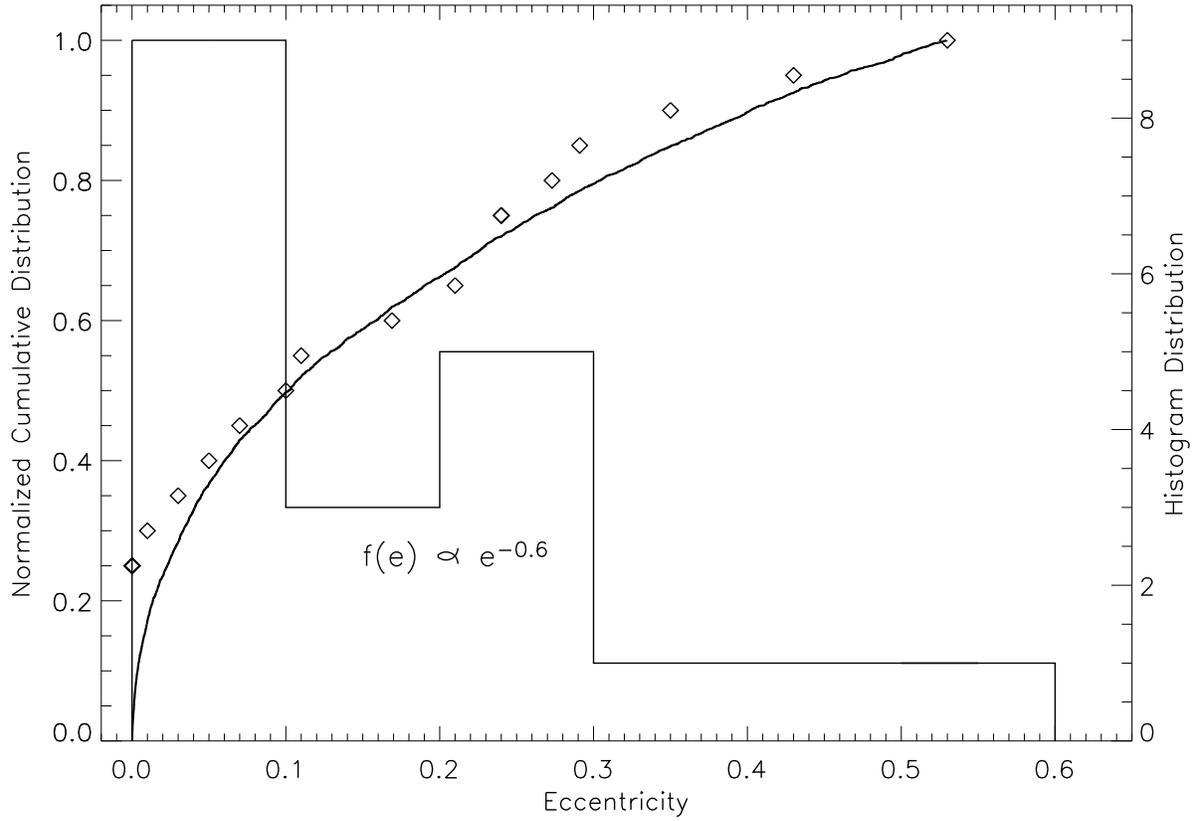}
\caption{Normalized cumulative distribution (diamonds) 
and conventional histogram for the
observed orbital eccentricities. The solid curve is the normalized cumulative
distribution for the best MC code fit, $\gamma=-0.6$. 
\label{edist}}
\end{figure}

\begin{figure}
\centering
\epsscale{1.0}
\plotone{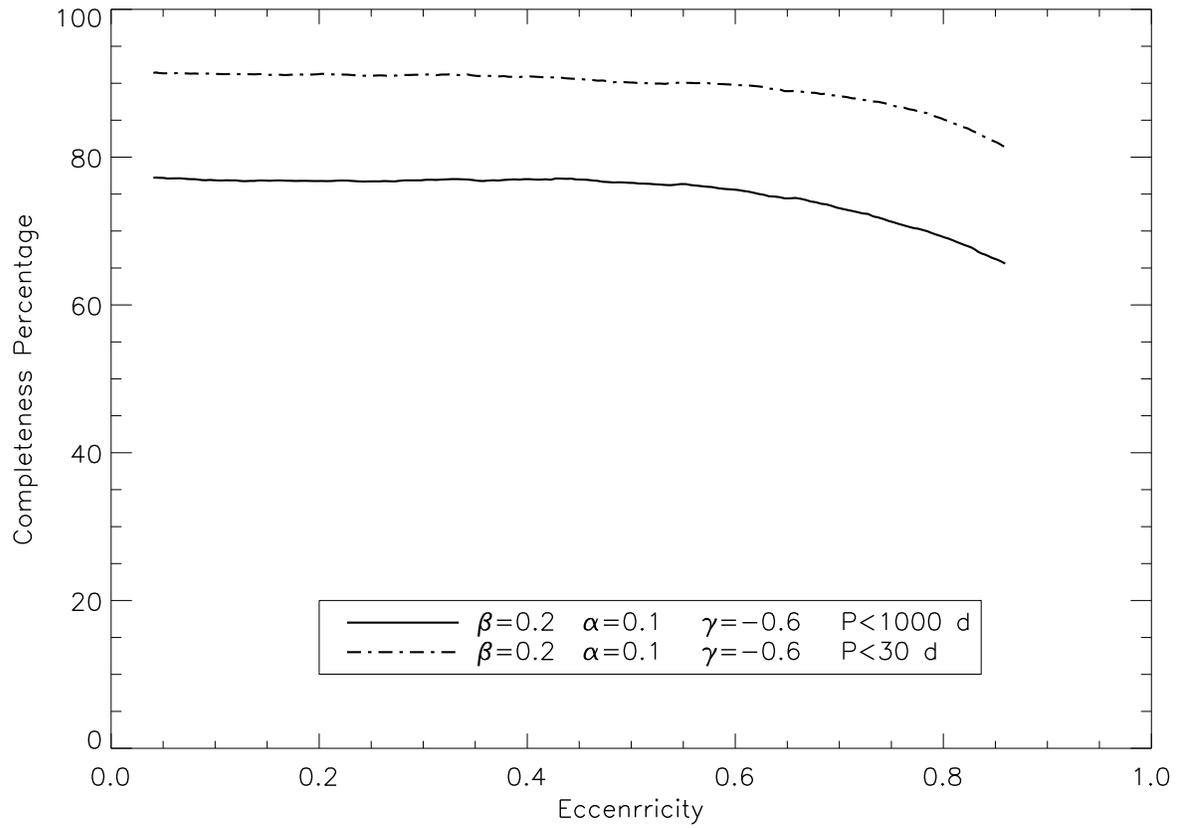}
\caption{Survey completeness as a function of eccentricity for the
best-fit power-law indices. The solid curve represents the period range
of 1--1000~days and the dot-dash curve represents the period range of
1--30~days.
\label{completeness_ecc}}
\end{figure}

\begin{figure}
\centering
\epsscale{1.0}
\plotone{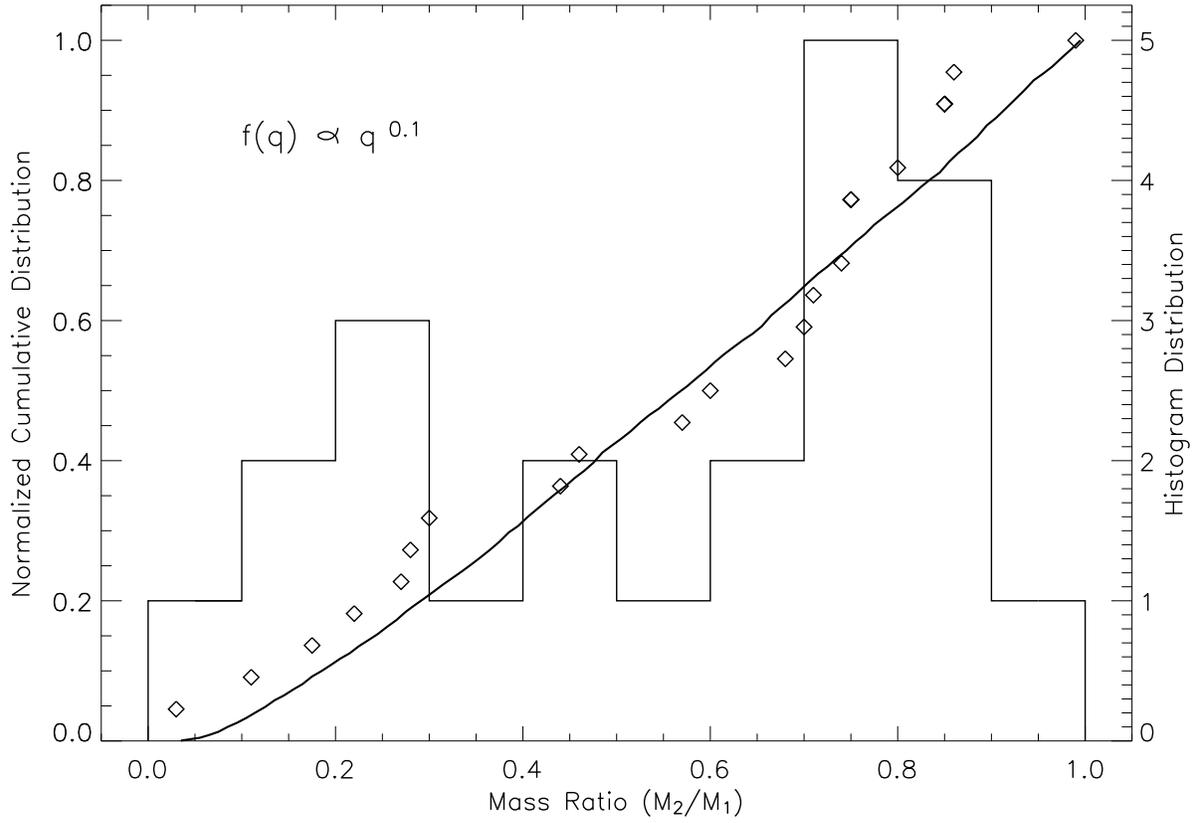}
\caption{Normalized cumulative distribution (diamonds) 
and conventional histogram for the
observed orbital mass ratios. The solid curve is the normalized cumulative
distribution for the best MC code fit, $\alpha=0.1$.
\label{qdist}}
\end{figure}

\begin{figure}
\centering
\epsscale{1.0}
\plotone{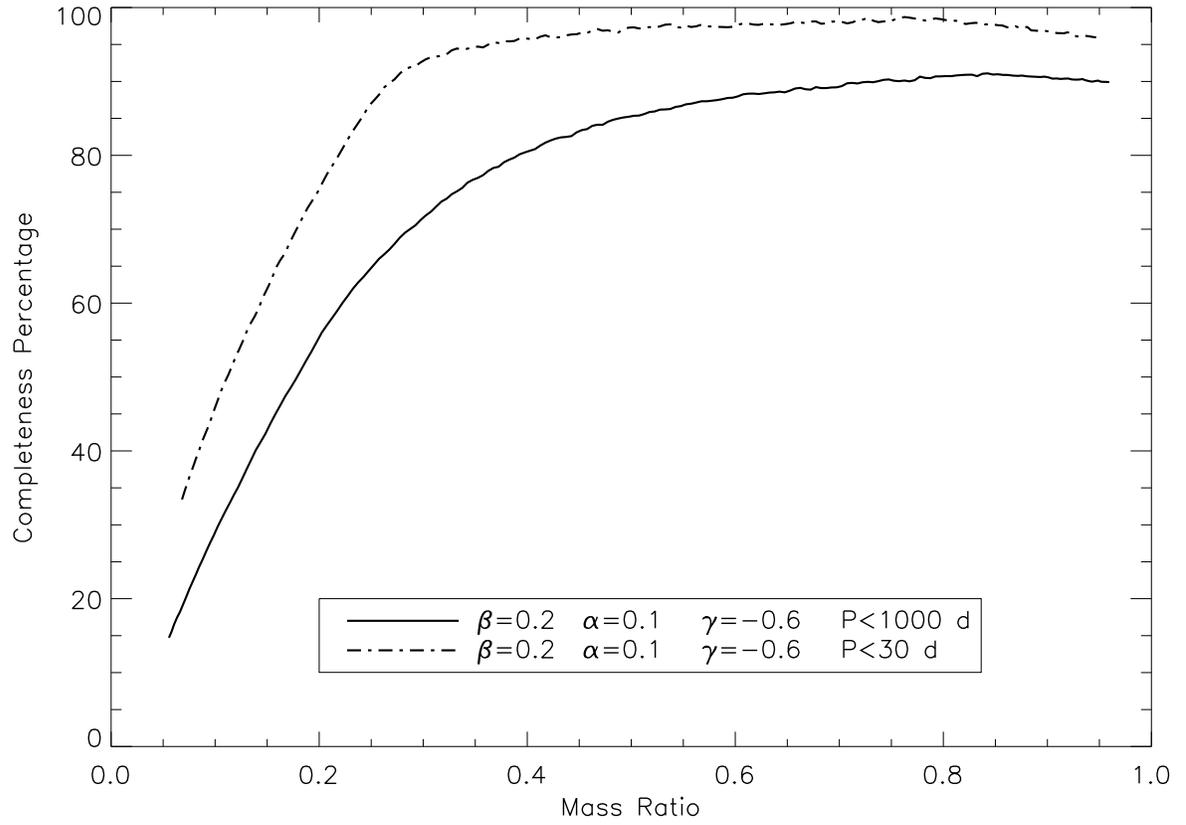}
\caption{Survey completeness as a function of mass ratio for systems with periods $P<30$ days (dot-dash curve) and $P<1000$ days (solid curve) for the adopted best-fit power-law indices.   
.
\label{completeness_mrs}}
\end{figure}

\begin{figure}
\centering
\epsscale{1.0}
\plotone{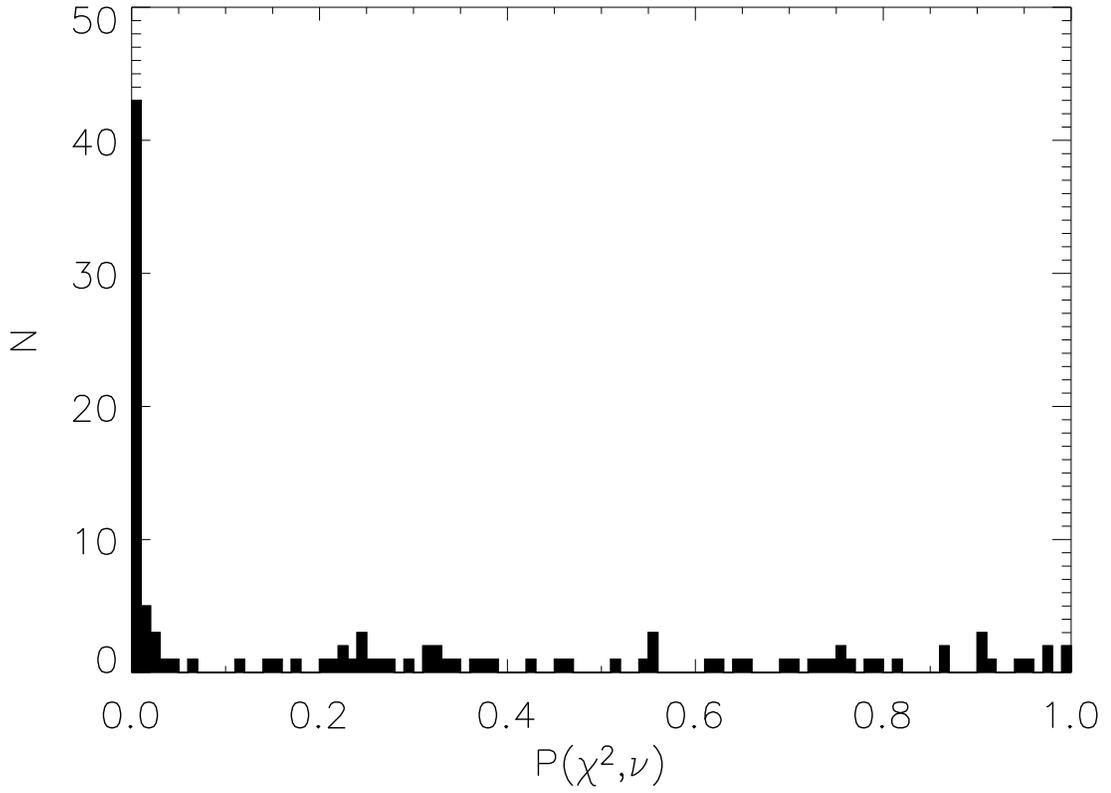}
\caption{$\chi^2$ probability density function for radial velocity 
measurements from the Cyg~OB2 survey.  
Systems with $P(\chi^2,\nu)<$5\%\ are probable velocity
variables.
\label{probchi}}
\end{figure}

\begin{figure}
\centering
\epsscale{1.0}
\plotone{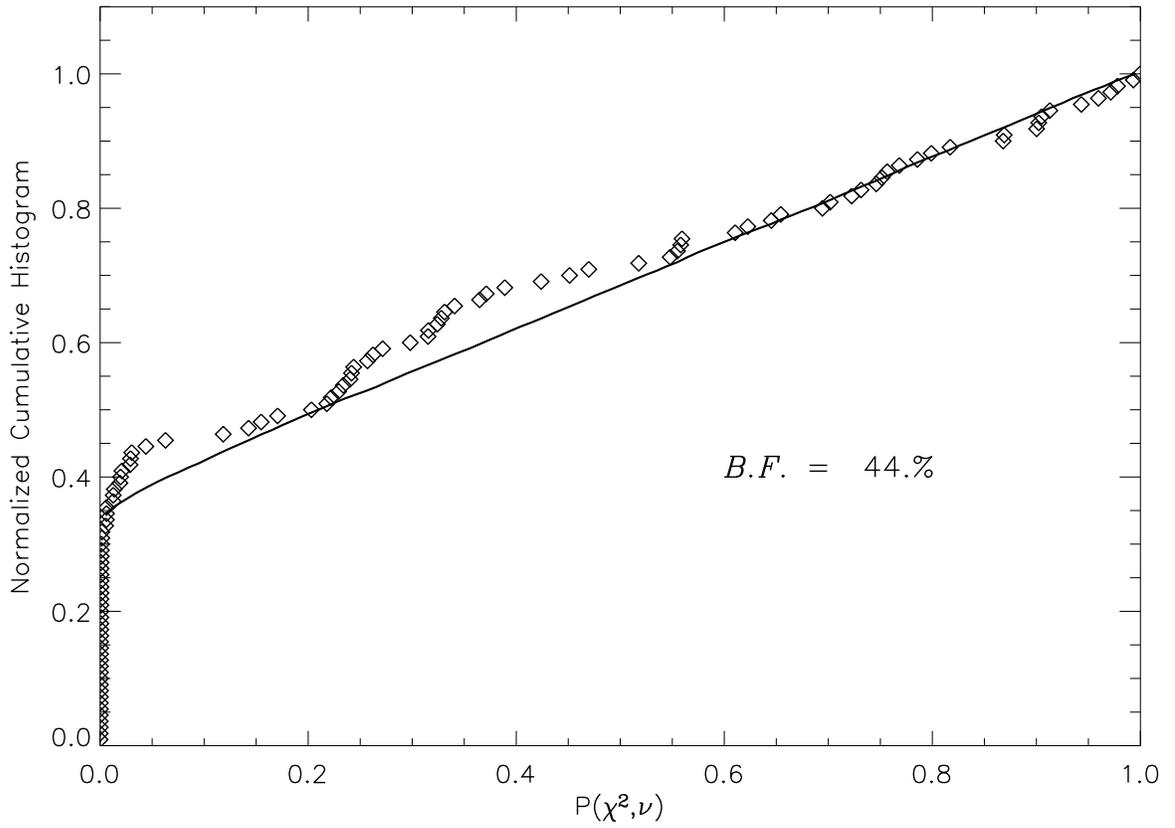}
\caption{The cumulative distribution of $\chi^2$ probabilities for the
observed data (diamonds) and the best MC code fit (solid curve). The
best fit represents a binary fraction of 44\%.
\label{chisqdist}}
\end{figure}

\begin{figure}
\centering
\epsscale{1.0}
\plotone{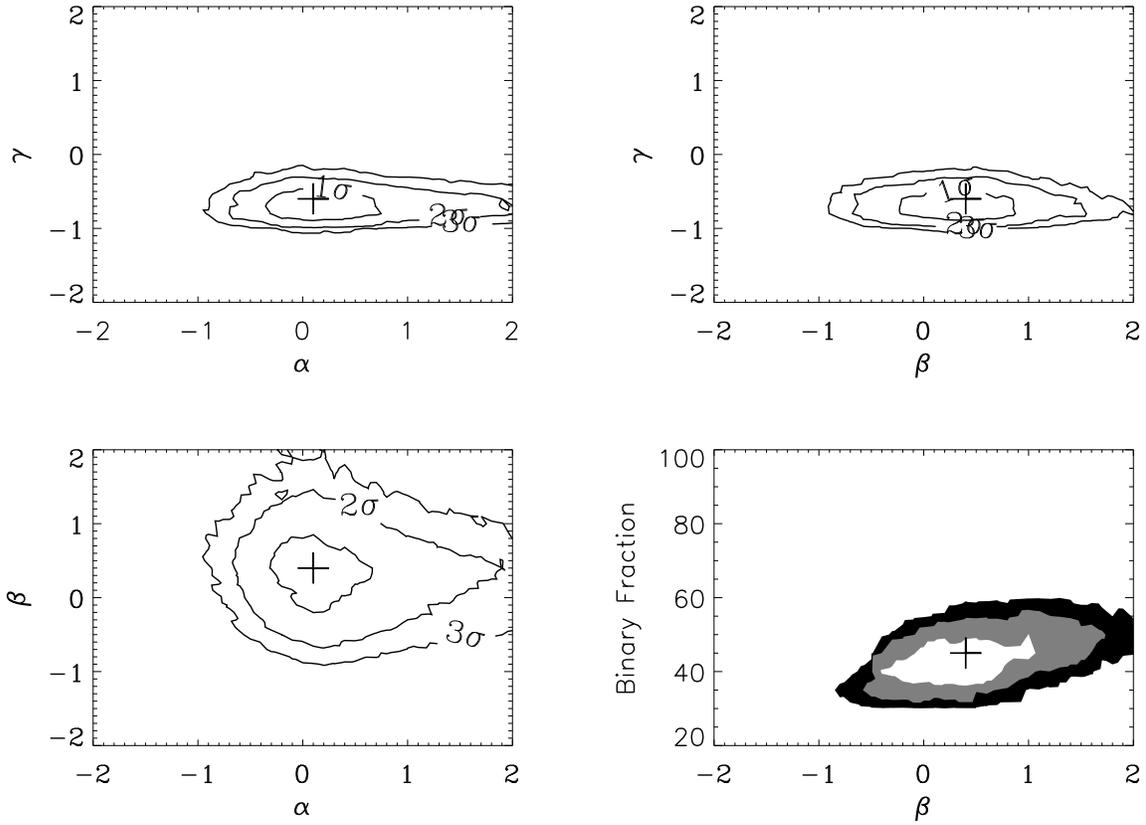}
\caption{Contour plots depicting the probabilities of K--S tests as in
Figure~\ref{Biased1}, but for the observed orbital parameter and
radial velocity data of the ``unbiased'' sample and the MC code fits
and with $B.F.=44$\%.
\label{Unbiased}}
\end{figure}

Figure~\ref{Unbiased} shows the results of the Monte Carlo analysis
for the 18 OB close binaries comprising the ``unbiased'' sample.
Similar to Figure~\ref{Biased1}, the four panels provide the
$1\sigma$, $2\sigma$, and $3\sigma$ probability contours for $\alpha$,
$\beta$, $\gamma$, and binary fraction.  The three panels depicting
probabilities for power-law indices (upper left, upper right, lower
left) are nearly identical to the results from the MC analysis of the
``complete'' sample shown in Figure~\ref{Biased1}.  The best-fit
$\beta$ is slightly larger (0.4 instead of 0.2) owing to the exclusion
of several shorter-period binaries from the unbiased sample.  The
best-fit $\alpha$ and $\gamma$ are the same, just with larger
uncertainties. The contours in the lower right panel show that the
binary fraction is confined to $37$\%~$<B.F.<49$\%\ at the $1\sigma$
level and $30$\%~$<B.F.<60$\%\ at the $3\sigma$ level for
$P_{MAX}=1000$ days.  This result is consistent with the range of
massive binary fractions for Galactic clusters listed in
Table~\ref{Ostat.tab}.  In order to compare our current results to the
Monte Carlo analysis of KF07, who explored a larger range of periods
(separations of 0.02--10,000 A.U. corresponding to
$P\sim$1--10$^{7}$~days for two 20 \msun\ stars), we also performed a MC
run with $P_{MAX}=10^7$ days.  In both analyses 10,000 A.U. is
considered a reasonable upper limit for bound systems.  In this case,
we find the binary fraction must be $\sim100$\%\ in order to replicate
our observed $\chi^2$ PDF. This is in rough agreement with the
conclusions of KF07, but based on extrapolating the power-law
distributions of mass ratios and periods determined from close
binaries over many orders of magnitude to characterize loosely bound
systems.

Is an extrapolation of the power law index $\beta$ to periods as long
as 10$^7$~days warranted when our radial velocity survey is sensitive
only to systems with periods less than about 10 years? \citet{Saida11}
conducted an adaptive optics and $HST$ Fine Guidance Sensor survey of
75 Cyg~OB2 systems and detect highly probable companions around 31 OB
primaries at separations $>$200 A.U. (P$>$630~years).  Of these,
nine (9/75 = 12\%) have probable periods in the range
$10^3$--$10^4$~years (S. Caballero-Nieves, private
communication)\footnote{We use a distance of 1700 pc, correct
statistically for projection effects, and assume a total system mass
of 30 M$_\odot$ to turn the angular separations into estimated
periods.}. If we adopt $P_{MAX}=10^4$ years ($\sim365,250$~days), the
MC code provides a best fit binary fraction of $\sim$90\%, and this
extrapolation of the $\beta\simeq 0.2$ power law predicts
approximately 15 binaries with orbital periods in each 1 dex interval
(i.e., 15 with periods between 10 yr and 100 yr, 15 between 100 yr and
1000 yr, etc.) given the 110 stars in our unbiased sample. This
translates into about 14\% of the binary systems falling within each
logarithmic period interval. Hence, the predictions of an extrapolated
power-law appear consistent with the emerging high angular resolution
observations at larger separations. We therefore suggest that the
binary fraction among massive stars may be as high as $90\pm10$\%\ if
periods as long as 10$^4$~years are allowed.

\subsection{Sensitivity to other model assumptions}

In addition to exploring the effect of varying $P_{MAX}$, we also
examined the effect of varying $q_{MIN}$ and the semi-amplitude
detection threshold, $K_{MIN}$, in the MC code.  As $K_{MIN}$ is
changed from 15 \kms\ to 20 \kms\ or larger, a greater fraction of
synthetic systems are deemed undetectable by the code, and the
detected systems have smaller separations and/or larger
companions. Therefore, we expect some combination of increase in
$\alpha$ (larger mass ratios), decrease in $\beta$ (shorter periods),
and increase in binary fraction. We find insignificant change (i.e.,
$\alpha$, $\beta$, $\gamma$ are the same) when $K_{MIN}=20$~\kms. A
significant change in the indices is not observed until $K_{MIN}$ is
increased to much larger values of $\sim50$~\kms. We are also
confident our detection threshold is much closer to 15~\kms, as
evidenced by the measurement of MT070 \citep{Kobulnicky2012}.  We
conclude that the minimum semi-amplitude for binary detection is
appropriately set at 15 \kms.

In the case of increasing $q_{MIN}$, the lowest mass ratio simulated
in the code, we primarily expect to see a decrease in $\alpha$ and a
decrease in the binary fraction. Binaries will have more massive
companions and semi-amplitudes will be larger, meaning that a smaller
fraction of binary systems will be deemed undetectable in the code.
We observed only a small decrease in binary fraction in runs with
$q_{MIN}=0.1$ (a few percent and still consistent with our initial
results). We also find only a small increase in $\alpha$ and no change
in $\beta$ or $\gamma$. However, increases in $\alpha$ become larger
when $q_{MIN}>0.1$.  For example, $\alpha$ changes from 0.1 to $-0.4$
for the ``complete'' sample when $q_{MIN}=0.2$; however, such a large
$q_{MIN}$ is probably unphysical given the existence of MT070
($q\ge0.03$) and MT267 ($q\ge0.11$).  These results, for both tests,
indicate that our MC results are not sensitive to small changes in the
adopted values of the mass ratio lower limit or semi-amplitude
threshold.

\section{Discussion}
\subsection{Possible Origin of the Short-Period Excess}

The presence of an excess of binary systems with periods around
3--5~days (with a peak at 4--5~days) and a shortage of systems with
periods around 7--14~days is not only observed in
Cyg~OB2. \citet{Sana08} observe a similar occurrence in NGC~6231, and
the excess and shortage of systems occur at roughly the same ranges in
period. Systems in Cyg~OB2 that lie within the excess encompass a
range of spectral types, mass ratios, and binary types (i.e.,
eclipsing, SB1 and SB2) and are scattered randomly across the cluster,
having no preferred location either in dense or sparse regions. We do
not believe this peculiar distribution can be attributed to
observational biases given the extended time coverage of this survey
and the different observing cadences of the observing runs that have
contributed to the Cyg~OB2 binary catalog.  While the existence of an
excess of systems with periods of around 3--5~days seems
well-established in the literature and in our present data, the
possibility of a shortage in the 7--14 day range has not been
previously noted.  A comparison between the Cyg~OB2 cumulative period
distribution with the data from \citet{garmany80} reveals a flattening
in this 7--14 day range (though we should be cautious given the use of
a standard histogram and small numbers). Their survey (of 28 systems)
contains three in this 7--14 day range while ours (22 systems)
contains one.

While we find that no single power-law can adequately describe the
log-period distribution in Cyg~OB2, it is unlikely, given any
power-law distribution, to find so few systems between 7 and 14 days.
We examined the probability that only a single system would be
observed in the 7--14 day period range given a random drawing of 22
systems with $1.46<P<25.2$~days (for $\beta=$-1 to 1). Random samples
were drawn 10,000 times and the final percentages were adjusted for a
90\%\ completeness level (the mean level as measured in
Figure~\ref{completeness_per} for the 7--14~day range). The likelihood
of detecting a single system for $\beta=0$ is $\sim$2\%. However, if
the power-law describing the log-period distribution is assumed to be
closer to $-1$, then the probability rises quickly to $\sim$13\%\
(with $\sim5$\%\ for $\beta=-0.5$, and $<1$\%\ for $\beta=$0.5 and 1).
This tells us that if the power-law(s) that describe the log-period
distribution have $\beta\geq-0.5$, the shortage is a real phenomenon
at the 2$\sigma$ level.

\begin{figure}
\centering
\epsscale{1.0}
\plotone{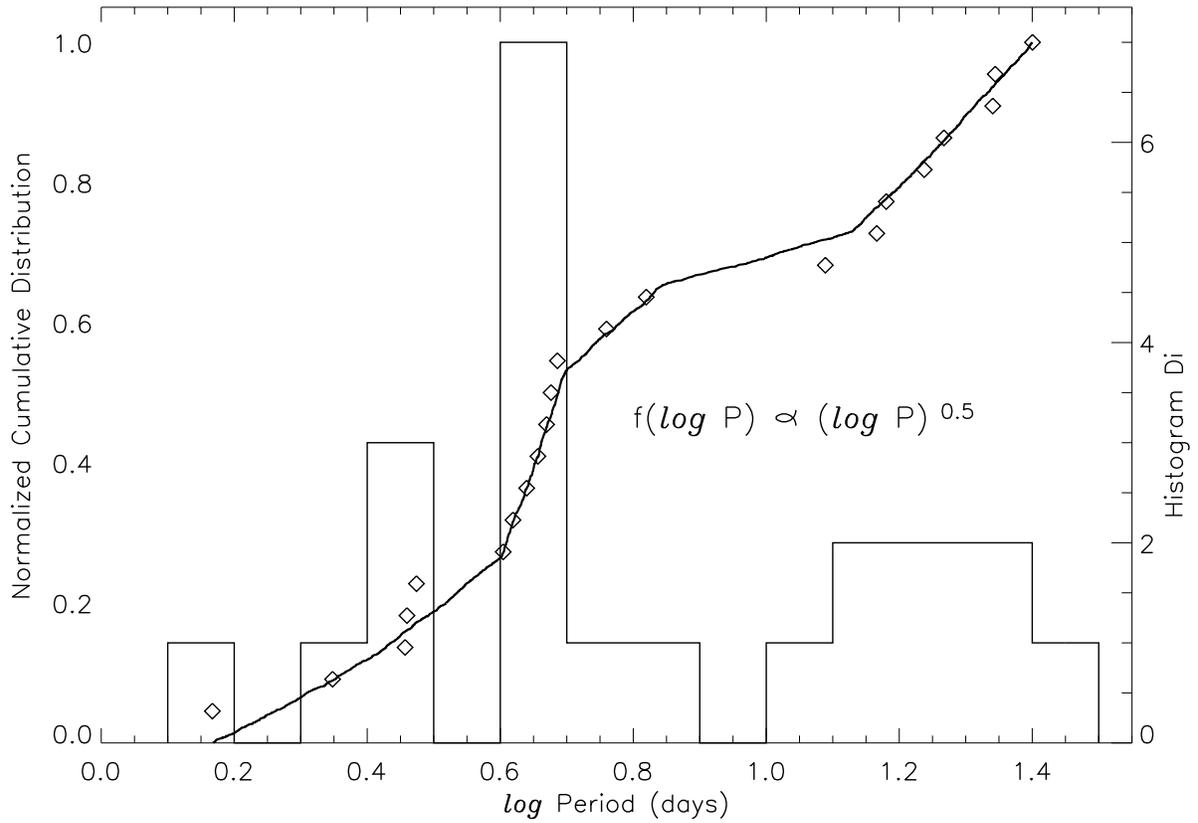}
\caption{Normalized cumulative distribution (diamonds) and
conventional histogram for the observed orbital periods. The solid
curve is the cumulative distribution for the best MC code fit,
assuming a 70\% orbital period migration from P=7--13.5~days to P=4--5~days.
\label{pdist_hump}}
\end{figure}

In an attempt to replicate the Cyg~OB2 period distribution with a
physically motivated prescription, we made a version of the Monte
Carlo code to take 70\%\ of the systems assigned a period between 7
and 14~days and reassign them assuming the same power-law describing
the rest of the distribution to periods between 3 and 5~days. The
70\%\ reassignment fraction reflects a 90\%\ completeness level and
detection of one out of a probable 3 or 4 systems in this range. This
approach approximates what would happen if binary systems with periods
7 and 14~days migrated to shorter periods. The best fit using this
approach is shown in Figure~\ref{pdist_hump} and comes from
reassigning systems with a period between 7 and 13.5~days to a period
between 4 and 5~days (corresponding to the peak of the excess in the
Cyg~OB2 distribution) with only a slight increase in $\beta$ (0.5 and
within the uncertainty of our results). $\alpha$, $\gamma$, and the
binary fraction are same as in Section~4. The normalized cumulative
distribution for the Cyg~OB2 data is represented by the diamonds and
the solid curve represents the normalized cumulative period
distribution for the best fit using this version of the MC
code. Though this fit requires a slightly more positive slope for the
initial period distribution, the overall fit to the data is visually
better than in Figure~\ref{pdist}, and the multiple changes in the
slope are also replicated well. A two-sided K--S test yields a
compatibility of greater than 99\%\ (compared to $\sim$60\%\ with the
simple power-law approach).

\begin{figure}
\centering
\epsscale{1.0}
\plotone{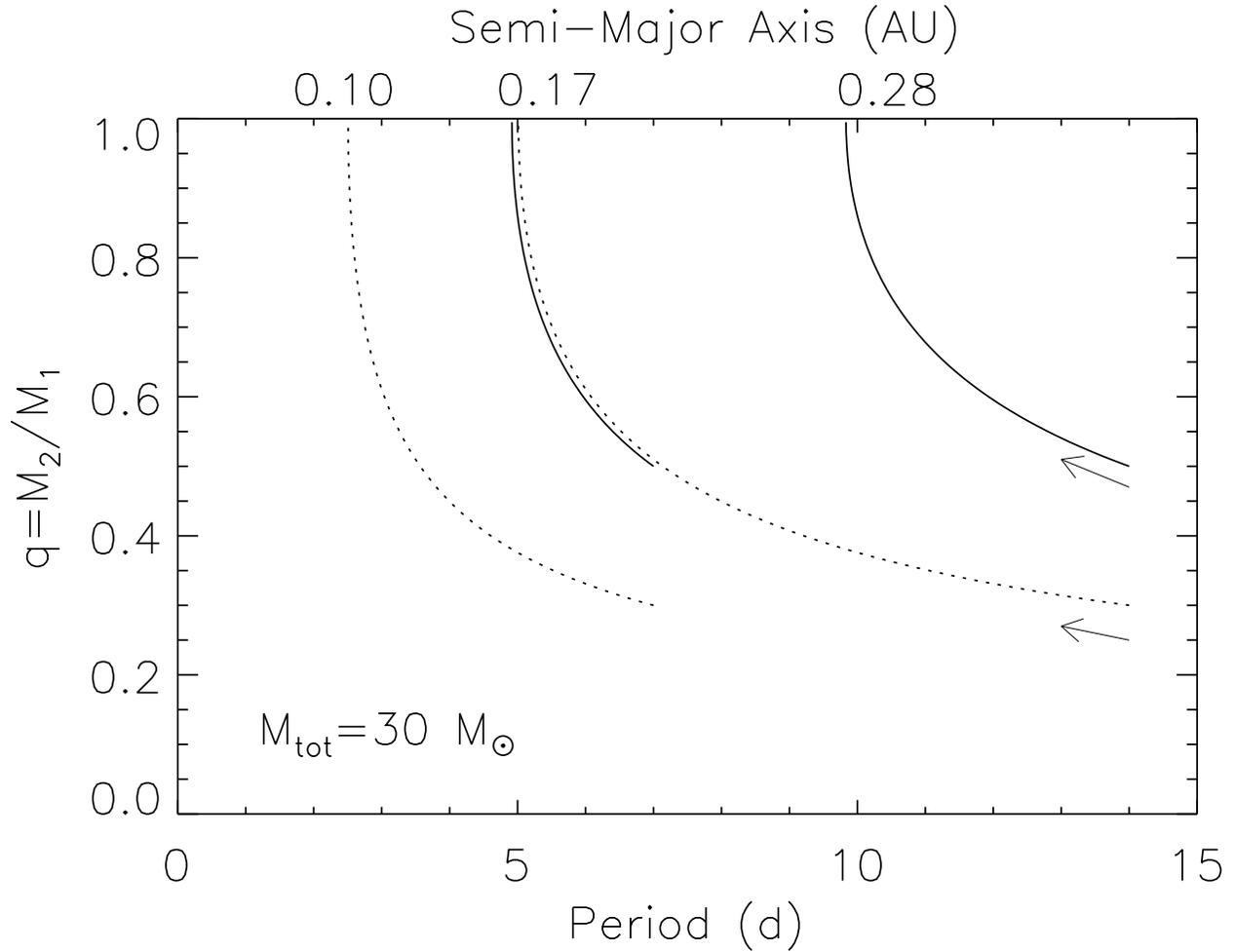}
\caption{The evolution of orbital period/semi-major axis and mass
ratio for a binary system with a total mass of 30~M$_\odot$ as the
primary loses mass to the secondary.  Arrows show the direction of
evolution as systems shift toward shorter periods (smaller orbits) and
larger mass ratio.  The solid and dotted tracks show the evolution of
systems with different initial masses ratios and orbital periods.
Systems with initial periods P=7--14 days and $q<$0.3 end up 
as systems near 3--5 days and $q\sim$1.
\label{pq30}}
\end{figure}

A plausible explanation for such period migration could be the
mechanism proposed by \citet{KrumThom07} whereby the circularization
of the orbit occurs primarily during the formation process instead. In
Figure~\ref{pq30} we show the evolution of orbital period/semi-major
axis and mass ratio for a binary system with a total mass of
30~M$_\odot$ as the primary loses mass to the secondary. The plotted
curves show tracks of constant angular momentum as systems evolve
toward shorter periods with mass ratios approaching unity, assuming that
mass and angular momentum are conserved,

\begin{equation}
{a \over a_0} = \left( {{q_0} \over {q}} \right)^2 \left( {{1+q} 
\over{1+q_0}}\right)^4 ,  
\end{equation}

\noindent (e.g., equation 8 of \citealt{KrumThom07}).  
Here $a_0$ and $q_0$ are the initial semi-major axis and mass ratio while
$a$ and $q$ are the final semi-major axis and mass ratio after mass transfer.
The two solid curves are tracks for systems
with initial mass ratios $q=0.5$ and initial periods of 7 and 14 days.
These systems may end up having mass ratios near unity, but their
final periods are longer than 5 days.  The dotted curves are tracks
for systems that begin with mass ratios $q=0.3$ and periods of 7 and
14 days.  These systems may end up with mass ratios near unity and
periods in the range 3--5 days.  This is precisely where we observe an
excess of systems with short periods in Cyg~OB2, and we also find a
relative abundance of systems with mass ratios $q=$0.8--0.9 --- not
exactly ``twins'', but systems where significant mass transfer may
have taken place.  This figure indicates that, if the
\citet{KrumThom07} mechanism is the operative one producing an
abundance of short-period, high-$q$ systems, then some fraction of the
population of 3--5 day systems should originate as systems with
$q\simeq0.3$ with 7--14 day periods.  While the evidence for
``missing'' 7--14 day systems in Cyg~OB2 is compelling, we cannot
claim, on the basis of the present data, to see the corresponding lack
of systems with $q\simeq0.3$.  Our survey is sufficiently incomplete
in this mass ratio range that observational biases preclude any firm
conclusion.

\subsection{Role of Close Binaries in Energetic Phenomena}

The Cyg~OB2 radial velocity survey originated as an observational
effort to inform theoretical models predicting the frequency of
energetic events like supernova and $\gamma$-ray bursts that may have
close binaries as progenitors. A key ingredient of these modeling
efforts is the fraction of massive stars that might interact through
mass transfer or undergo a common envelope phase, especially during
the post-main-sequence evolution of either star in the system. Red
supergiants are the descendants of massive stars, so we adopt the RSG
radii deduced by \citet{levesque05} which range between 200 R$_\odot$
and 1500 R$_\odot$.  Their Figure~3 shows stellar evolutionary tracks
indicating that stars in the mass range 15 -- 30 M$_\odot$ may reach
or exceed 1000 R$_\odot$ while stars with initial masses over 10
M$_\odot$ can generally be expected to reach 500 R$_\odot$.  We adopt
500 R$_\odot$ as a maximum evolved radius for stars in our sample,
meaning that systems with semi-major axes smaller than this value will
interact strongly, probably through the formation of a common
envelope.  Other systems may interact via mass transfer after Roche
Lobe overflow.  \citet{Eggleton83} provides an analytic description of
the Roche Lobe radius as a function of mass ratio.  His tabulated
values show that the Roche Lobe radii for binary systems with mass
ratios $q= $0.2 -- 1.0 fall in the range 0.38$a$ -- 0.55$a$ where $a$
is the separation of the components.  Consequently, we adopt 1000
$R_\odot$ (4.7 A.U.)  as a conservative threshold separation, below
which systems are deemed to interact during the evolution of the
system. In our Monte Carlo code, we flag all systems with periastron
distances less than 1000 $R_\odot$ as capable of interaction. Assuming
$\alpha=0.1$, $\beta=0.2$, $\gamma=-0.6$, and $P_{MAX}=10^4$~years,
this predicts that 45\%\ of the binary systems will have sufficiently
close companions to interact during the binary's lifetime (37\%\ if we
adopt the slightly larger $\beta=0.5$ from Section~5.1).

\subsection{The Luminosity Contribution from Secondaries}

The luminous contribution from secondary stars, which are
unresolved in photometric surveys, even with the $Hubble Space
Telescope$ or adaptive optics, is not negligible.  Consequently,
measuring accurate distances to star clusters using spectroscopic
parallax, main sequence fitting, or similar standard-candle techniques
is systematically in error if close companions are neglected.  As part
of our Monte Carlo code we tallied the mean V-band light contribution
from primary and secondary stars by using a zero-age main sequence
(ZAMS) relation between stellar mass and absolute V magnitude
\citep{FM05, drilling}.  For the best fitting power law indices
described above, allowing for periods as long as 10$^4$ yr, we find
that the secondary stars contribute 16\% of the total stellar light at
V-band.  This is equivalent to a systematic error in the distance
modulus of 0.16 mag and a systematic distance error of about 8\%, in
the sense that distances derived without correcting for luminous
secondaries will be too small.  We expect the secondary luminosity
fractions in evolved populations to vary stochastically about this
16\% ZAMS value as the primary and secondary components throughout the
population become giants and supergiants. The intricacies of mass
exchange between components will further limit any simple description
of the time evolution in the nominal ZAMS value.

\section{Conclusions}

In this, our fifth work of this series, we present the results of a
Monte Carlo analysis using a decade of radial velocity data on 114
massive stars in the Cyg~OB2 Association.  This survey is unique in
terms of number of massive stars and binary solutions analyzed. It
improves on KF07 and other works (e.g., \citealt{Evans06},
\citealt{garmany80}, \citealt{Gies87}, \citealt{Pour04}) in that we
utilize the radial velocity information obtained over a larger time
span and make use of the nearly complete orbital
information---including eccentricities---on 22 of the 24 known
massive binaries.  The Cyg~OB2 binaries have been observed by
spectroscopic and/or photometeric means and cover a period range of
1.47~days -- 7.58~years (with all but a few having $P<26$~days), an
eccentricity range of 0--0.53, and a mass ratio range of 0.03--0.99. 
On the strength of these data, we compute the binary fraction for a
large, young population (2--3~Myr) of massive stars from a common
origin.  In addition, we use this information to model the most
probable intrinsic distributions for period, mass ratio, and
eccentricity for short-period systems having P$<$26 days.  

\begin{itemize}

\item{The data imply a hard lower limit on the binary fraction of 21\%
(i.e., 24/114 systems).  The Monte Carlo analysis of the unbiased
sample of 110 OB stars (i.e., not including A36, A45, B17, and
Schulte~5), indicates a massive star binary fraction of $44\pm8$\%,
assuming that the allowed period range is 1--1000 days (separations
$\sim$0.01 -- few $\times$ 10~AU).  However, if we extrapolate the
best fit power-law distributions over several orders of magnitude and
allow for periods as long as 10$^4$ years, the binary fraction may be
as large as $90\pm10$\%. These results are consistent with the
probable number of binaries detected in the adaptive optics and
\textit{HST} survey of Cyg~OB2 by \citet{Saida11} and with the
previous results of KF07.}

\item{ The best-fitting distributions of mass ratio, period, and
eccentricity are described by power-laws with indices
$\alpha=0.1\pm0.5$, $\beta=0.2\pm0.4$, and $\gamma=-0.6\pm0.3$,
respectively, assuming lower and upper bounds on the power-law
distributions of q=0.005--1.0, P=1--1000~days, and e=0.0001--0.9.
These values for $\alpha$ and $\beta$ are broadly consistent with the
previous results of KF07, as well as \citet{Sana08},
\citet{Kouwenhoven07}, \citet{garmany80}, and the analysis of
\citet{Sana10a}. The mass ratio distribution is not consistent with
the ``twin-heavy'' population proposed by \citet{pinson06}, but
suggests a weak preference for larger mass ratios (i.e., $q\ge0.8$). }

\item{The distribution of orbital periods is not consistent with any
single power-law, owing to a significant excess of systems with
periods around 3--5~days (peaking at 4--5~days) and a deficit of
systems with periods around 7--14 days.  Such a 3--5 day excess has
been observed among massive stars in NGC~6231 \citep{Sana08}, but the
latter phenomenon has not been previously reported. We propose that it
may stem from binary systems with periods initially in the range of
$\sim$7--14~days migrating to shorter periods during a period of
pre-main-sequence mass transfer, via the mechanism proposed by
\citet{KrumThom07}.}

\item{The fraction of massive-star binaries (or higher order systems)
 having companions close enough to interact during pre- or
 post-main-sequence evolution, defined here as requiring a semi-major
 axis $<$4.7~AU ($\sim$1000~\rsun), is $\sim$45\%. This fraction is a
 key ingredient in recipes that predict the cosmic rates of
 binary-induced phenomena arising from massive stars, such as some
 classes of supernova and $\gamma$-ray bursts.  }

\end{itemize}

We present these findings in the anticipation that they, along with
the analyses of other clusters (such as the work with Sco OB2 by
\citealt{Kouwenhoven07}, NGC~6231 by \citealt{Sana08}, NGC~2244 by
\citealt{Mahy09}, and NGC~6611 by \citealt{Sana09}), will aid our
understanding of numerous issues involving massive stars, including
their formation, the predicted rates of supernovae type Ib/c arising
through binary channels, and whether there is a correlation between
cluster density and the binary fraction therein.  This ongoing survey
will continue to uncover massive spectroscopic binaries in Cyg~OB2.
These new SB1s and SB2s will help refine our characterization of the
orbital parameter distributions and increase our understanding of the
nature of the excess in orbital periods between $\sim$3--5~days.
While the present data has provided solid constraints on the massive
binary characteristics in Cyg~OB2, it is possible that the binary
fraction or distribution of periods/mass ratios/eccentricities varies
temporally as a cluster ages or is a function of initial cluster
density.  Discerning whether these additional variables are important
will require more extensive surveys over a range of age and formation
environment.

\acknowledgements We would like to acknowledge the helpful advice and
attention to detail we received from our referee in the process of
making this a much better manuscript. We thank Mark Krumholz and Hans
Zinnecker for inspiring conversations about phenomena in short-period
binaries, Chris Fryer for helpful discussions regarding interacting
binary systems, and Saida Caballero-Nieves for sharing her imaging
data of Cyg~OB2 in advance of publication.  We are very appreciative
for the continued advice and encouragement throughout the project from
Ginny McSwain. Additionally, we are grateful for the continued support
from the National Science Foundation through Research Experience for
Undergraduates (REU) program grant AST 03-53760, through grant AST
03-07778, AST 09-08239, and the support of the Wyoming NASA Space
Grant Consortium through grant NNG05G165H.

\textit{Facilities:} \facility{WIRO ()}, \facility{WIYN ()},
\facility{Shane ()}, \facility{Keck:I ()}

\clearpage

\begin{deluxetable}{lrrr}
\tabletypesize{\scriptsize}
\tablecaption{Ephemerides for Cyg OB2 Stars Used in Monte Carlo Analysis \label{XX.tab}}
\tablewidth{0pt}
\tablehead{
\colhead{} &
\colhead{Date} &
\colhead{$V_{r}$} &
\colhead{$\sigma_V$} \\
\colhead{Star} &
\colhead{HJD-2,400,000} &
\colhead{(\kms)} &
\colhead{(\kms)}}
\startdata
A36 & 54,403.61 &  -163.4 &     5.6 \\ 
A36 & 54,403.70 &  -174.9 &     5.0 \\ 
A36 & 54,405.71 &   103.4 &     5.3 \\ 
A36 & 54,406.63 &   105.4 &     3.7 \\ 
A36 & 54,406.76 &   108.6 &     3.7 \\ 
\enddata
\tablecomments{Column~1 gives the star designation in either
\citet{Schulte58} (S), \citet{Comeron02} (A,B) or \citet{MT91} (MT) 
format. Column~2
gives the heliocentric Julian date in the format HJD--2,400,000.
Column~3 gives the cross-correlation velocity in \kms. Column~4 gives
the 1$\sigma$ cross-correlation error in \kms. Table~1 is published in
its entirety in the electronic edition of the Astrophysical Journal. A
portion is shown here for guidance regarding its form and content.}
\end{deluxetable}

\clearpage

\begin{deluxetable}{lrrrrr}
\centering
\tabletypesize{\scriptsize}
\tablecaption{Massive Star Binary Statistics \label{Ostat.tab}}
\tablewidth{0pt}
\tablehead{
\colhead{Cluster} &
\colhead{O Stars in} &
\colhead{Early-B Stars} &
\colhead{$b.f.$} &
\colhead{$b.f.$} &
\colhead{Reference} \\
\colhead{ID} &
\colhead{Survey} &
\colhead{in Survey} &
\colhead{(O-Stars)} &
\colhead{(Massive Stars)} &
\colhead{ID}}   
\startdata
NGC 2004 (LMC)   & 4 (1)    & 83 (21)  & 25\%\                      & 25\%\    & 1      \\
NGC 330 (SMC)    & 6 (0)    & 76 (3)   & 0\%\                       & 4\%\     & 1      \\
NGC 346 (SMC)    & 19 (4)   & 59 (19)  & 21\%\                      & 29\%\    & 1      \\
Trumpler 14      & 7 (1)    & \nodata  & 14\%\                      & \nodata  & 2,3    \\
IC 1805          & 10 (2)   & \nodata  & 20\%\                      & \nodata  & 4,5    \\
NGC 2244         & 6 (1--2) & \nodata  & 17--33\%\                  & \nodata  & 6      \\
NCC 6231         & 16 (10)  & 33 (16)  & 63\%\                      & 53\%\    & 2,7    \\
NGC 6611         & 9 (4)    & \nodata  & 44--67\%\tablenotemark{a}  & \nodata  & 2,3,8  \\
IC 2944          & 16 (7)   & \nodata  & 44\%\                      & \nodata  & 2,3    \\
Trumpler 16      & 20 (7)   & \nodata  & 35\%\                      & \nodata  & 2,3    \\
Cr 228           & 21 (5)   & \nodata  & 24\%\                      & \nodata  & 2,3    \\
Cyg~OB2          & 44 (12)  & 69 (8)   & 27\%\                      & 18\%\    & 9,10,11 \\
\enddata
\tablecomments{A selection of OB cluster/association studies examining
the binary fraction of massive stars. The number of binaries are shown
in parentheses.}
\tablerefs{(1) \citet{Evans06}; (2) \citet{Garcia01}; (3) \citet{Merm95}; 
(4) \citet{debeck06}; (5) \citet{Rauw04}; (6) \citet{Mahy09}; 
(7) \citet{Sana08}; (8) \citet{Sana09}; (9) \citet{Kiminki08}; 
(10) \citet{Kiminki09}; (11) \citet{Kiminki2010}}

\tablenotetext{a}{Upper bound obtained from modeling complete fraction.}

\end{deluxetable}

\clearpage

\clearpage

\clearpage

\clearpage

\clearpage

\clearpage

\clearpage

\clearpage

\clearpage

\clearpage

\clearpage

\clearpage

\clearpage

\clearpage

\end{document}